\documentclass[11pt]{article}
\usepackage{amsmath}
\usepackage{hyperref}
\usepackage{wasysym}
\usepackage[font=footnotesize]{caption}
\usepackage[T1]{fontenc}
\usepackage{latexsym}
\usepackage[english]{babel}
\usepackage{cite}
\usepackage{enumerate}
\usepackage[shortlabels]{enumitem}
\pdfoutput=1
\usepackage[utf8]{inputenc}
\usepackage{dsfont}
\usepackage{diagbox}
\usepackage{amsfonts}
\usepackage{mathtools}
\allowdisplaybreaks[4]        
\usepackage{amssymb}
\usepackage{euscript}     
\usepackage{braket}
\usepackage{amssymb}
\usepackage{starfont}
\usepackage{color,soul}         
\usepackage{braket}
\usepackage{tensor}        
\usepackage{amsthm}
\usepackage{graphicx}
\usepackage{slashed}
\usepackage{leftidx}
\usepackage{subfigure}
\usepackage{bbm}
\usepackage{empheq}
\usepackage[dvipsnames]{xcolor}
\definecolor{outerspace}{rgb}{0.25, 0.29, 0.3}
\definecolor{scarlet}{rgb}{1.0, 0.13, 0.0}
\usepackage[header,title,page,titletoc]{appendix}  
\definecolor{princetonorange}{rgb}{1.0, 0.56, 0.0}
\definecolor{WildStrawberry}{rgb}{1.0, 0.26, 0.64}
\definecolor{rossocorsa}{rgb}{0.83, 0.0, 0.0}
\definecolor{navyblue}{rgb}{0.0, 0.0, 0.5}
\usepackage[numbers,sort&compress]{natbib}  
\usepackage{float}
\usepackage[paper=a4paper,margin=1in]{geometry}
\usepackage{titlesec}

\pdfoutput=1
\usepackage{amsmath}
\usepackage{color}
\usepackage{amsfonts}
\usepackage{amssymb}
\usepackage{graphicx}
\usepackage{geometry}
\usepackage{amssymb,epsfig,subfigure}
\usepackage{amssymb}
\usepackage{hyperref}
\usepackage{comment}
\usepackage[font=footnotesize]{caption}
\usepackage[T1]{fontenc}
\usepackage[utf8]{inputenc}
\usepackage{latexsym}

\usepackage{cancel}

\usepackage[numbers,sort&compress]{natbib}  
\usepackage{float}

\usepackage{dsfont}
\usepackage{diagbox}

\usepackage{mathtools}
\allowdisplaybreaks[4]

\usepackage{euscript}     
\usepackage{braket}
\usepackage{starfont}
\usepackage{color,soul}         
\usepackage{tensor}        
\usepackage{amsthm}

\usepackage{slashed}
\usepackage{leftidx}
\usepackage{bbm}

\makeatletter
\renewcommand\section{\@startsection {section}{1}{\z@}%
                                 {-3.5ex \@plus -1ex \@minus -.2ex}
                                   {2.3ex \@plus.2ex}%
                                   {\normalfont\large\bfseries}}
\renewcommand\subsection{\@startsection{subsection}{2}{\z@}%
                                   {-3.25ex\@plus -1ex \@minus -.2ex}%
                                     {1.5ex \@plus .2ex}%
                                     {\normalfont\bfseries}}
\renewcommand\subsubsection{\@startsection{subsubsection}{3}{\z@}%
                                   {-3.25ex\@plus -1ex \@minus -.2ex}%
                                     {1.5ex \@plus .2ex}%
                                     {\normalfont\itshape}}
\makeatother

\def\pplogo{\vbox{\kern-\headheight\kern -29pt
\halign{##&##\hfil\cr&{\ppnumber}\cr\rule{0pt}{2.5ex}&\ppdate\cr}}}
\makeatletter
\def\ps@firstpage{\ps@empty \def\@oddhead{\hss\pplogo}%
  \let\@evenhead\@oddhead 
}
\def\maketitle{\par
 \begingroup
 \def\thefootnote{\fnsymbol{footnote}}
 \def\@makefnmark{\hbox{$^{\@thefnmark}$\hss}}
 \if@twocolumn
 \twocolumn[\@maketitle]
 \else \newpage
 \global\@topnum\z@ \@maketitle \fi\thispagestyle{firstpage}\@thanks
 \endgroup
 \setcounter{footnote}{0}
 \let\maketitle\relax
 \let\@maketitle\relax
 \gdef\@thanks{}\gdef\@author{}\gdef\@title{}\let\thanks\relax}
\makeatother

\numberwithin{equation}{section}

\newcommand\eea{\end{eqnarray}}
\newcommand\bea{\begin{eqnarray}}

\def\beq{\begin{equation}}
\def\eeq{\end{equation}}

\newcommand{\be}{\begin{equation}}
\newcommand{\ee}{\end{equation}}
\newcommand{\ba}{\begin{align}}
\newcommand{\ea}{\end{align}}
\newcommand{\bg}{\begin{gather}}
\newcommand{\eg}{\end{gather}}
\newcommand{\bseq}{\begin{subequations}}
\newcommand{\eseq}{\end{subequations}}

\usepackage{hyperref}
\hypersetup{
    colorlinks,
    citecolor=rossocorsa,
    filecolor=navyblue,
    linkcolor=navyblue,
    urlcolor=navyblue
}

\begin{document} 

\begin{titlepage}

\begin{center}

\phantom{ }
\vspace{3cm}

{\bf \Large{Generalized symmetries and Noether's theorem in QFT}}
\vskip 0.5cm
Valentin Benedetti${}^{\dagger}$, Horacio Casini${}^{*}$, Javier M. Mag\'an${}^{\ddagger}$
\vskip 0.05in
\small{ ${}^{\dagger}$ ${}^{*}$ \textit{Instituto Balseiro, Centro At\'omico Bariloche}}
\vskip -.4cm
\small{\textit{ 8400-S.C. de Bariloche, R\'io Negro, Argentina}}
\vskip -.10cm
\small{${}^{\ddagger}$ \textit{David Rittenhouse Laboratory, University of Pennsylvania}}
\vskip -.4cm
\small{\textit{ 209 S.33rd Street, Philadelphia, PA 19104, USA}}
\vskip -.10cm
\small{${}^{\ddagger}$ \textit{Theoretische Natuurkunde, Vrije Universiteit Brussel (VUB) }}
\vskip -.4cm
\small{\textit{ and The International Solvay Institutes}}
\vskip -.4cm
\small{\textit{ Pleinlaan 2, 1050 Brussels, Belgium}}

\begin{abstract}
We show that generalized symmetries cannot be charged under a continuous global symmetry having a Noether current. Further, only non-compact generalized symmetries can be charged under a continuous global symmetry. These results follow from a finer classification of twist operators, which naturally extends to finite group global symmetries. They unravel topological obstructions to the strong version of Noether's theorem in QFT, even if under general conditions a global symmetry can be implemented locally by twist operators (weak version). We use these results to rederive  Weinberg-Witten's theorem within local QFT, generalizing it to massless particles in arbitrary dimensions and representations of the Lorentz group. Several examples with local twists but without Noether currents are described. We end up discussing the conditions for the strong version to hold, dynamical aspects of QFT's with non-compact generalized symmetries, scale vs conformal invariance in QFT, connections with the Coleman-Mandula theorem and aspects of global symmetries in quantum gravity.
\end{abstract}
\end{center}

\small{\vspace{3.5 cm}\noindent ${}^{\dagger}$valentin.benedetti@ib.edu.ar\\
${}^{\text{\text{*}}}$casini@cab.cnea.gov.ar\\
${}^{\ddagger}$ magan@sas.upenn.edu
}

\end{titlepage}

\setcounter{tocdepth}{2}

{\parskip = .4\baselineskip \tableofcontents}
\newpage


\section{Introduction}
\label{uno}

In classical mechanics,  Noether's theorem \cite{Noether1918} asserts the existence of conserved charges $Q$ when the action is invariant under a continuous symmetry group. These conserved charges are the generators of the group in the classical phase space of theory. An a priori stronger version appears in classical field theory, where Noether's theorem implies the existence of local conserved currents $j^{\mu}$, satisfying $\partial_{\mu}\,j^{\mu}=0$. By integrating these currents over a space-like surface, one obtains the conserved charges. 

It is a long-standing question to determine to what extent, or under what conditions, this theorem holds in Quantum Field Theory (QFT). In most scenarios,  this is known to be the case, and the existence of a global continuous symmetry entails the existence of a local conserved current. The appearance of local currents associated with global continuous symmetries will be called ``the strong version'' of Noether's theorem.

In fact, there is a weaker version of Noether's theorem. If we have a current,  we can integrate it over a finite part $R$ of the full space-like surface. This partial integration gives rise to the local charges $Q_R$, which measure the amount of charge in the region $R$. Equivalently, they are the generators of localized symmetry transformations for local fields in $R$. These localized symmetry transformations are called twists. The existence of local charges $Q_R$ and twists is ensured by the existence of the currents themselves. But the validity of the converse statement is unknown. The existence of local $Q_R$ for any region $R$ will be called ``the weak version'' of Noether's theorem.

Significant progress towards a Lagrangian independent, first principles proof of Noether's theorem in QFT appeared in Ref. \cite{doplicher1982local,Doplicher:1983if,Doplicher:1984zz,buchholz1986noether}, within the context of the algebraic approach to global symmetries in QFT \cite{Doplicher:1969tk,Doplicher:1969kp,doplicher1971local,doplicher1974local,Doplicher:1990pn}. In particular, in Ref. \cite{buchholz1986noether}, given a certain mild technical assumption called the split property in QFT \cite{Doplicher:1984zz}, the weak version of Noether's theorem was proven, and a canonical construction of the local charges $Q_R$ was found. We review this construction below. It is also valid for discrete global symmetries, extending the weak version of the theorem to those cases as well. Finally, Ref. \cite{buchholz1986noether} also acknowledges that this construction is not a proof of the strong version of Noether's theorem, and that further new ideas are required to accomplish such a goal, or to find the potential obstruction.  Progress in this direction can be found in \cite{Current1,Current2,Harlow:2018tng}.

An a priori unrelated theorem in 4d QFT is the Weinberg-Witten theorem \cite{WEINBERG198059}. This theorem states that the existence of a conserved current $j^\mu$ precludes the existence of massless particles with helicity greater or equal to one that are charged under it, and that the existence of a stress tensor $T^{\mu\nu}$ precludes the existence of massless particles with helicity greater or equal to two. These theorems then constrain the space of possible QFT's in 4d. In particular, the second version is typically interpreted as an obstruction to constructing relativistic QFT's with an emergent graviton in the infrared, suggesting that the quantization of gravity must follow a different route. 

It is plain that both theorems, the strong version of Nother's theorem and the Weinberg-Witten theorem, are calling for a deeper understanding of when and why conserved currents generating global symmetries can appear in the theory. An observation suggesting that both of them are related to each other appears when turning around the Weinberg-Witten theorem. For example, if a theory of massless particles with helicity greater or equal to one in 4d charged under a global symmetry exists, it would definitely violate the strong version of Noether's theorem since according to Weinberg-Witten there is no local conserved current for such a symmetry.

Given this context, the main motivation of this article is to understand the space of QFT's violating the strong version of Noether's theorem, and analyze the lessons learned in relation to the Weinberg-Witten theorem. To such end,  we develop a finer classification of localized charges and twist operators in QFT. This classification follows naturally from the local QFT approach to generalized symmetries developed in \cite{Casini:2020rgj,Review}, which we review in the next section. More precisely, to make progress it will be crucial to consider QFT's with generalized symmetries on top of the global symmetry in which we are focusing our attention. In such scenarios, the associated charges/twists in a certain region $R$ potentially come in different types, which we analyze in detail.

Using this finer classification of twist operators we derive the main results of this article, from where the others follow. In words, we show that generalized symmetries cannot be charged under a continuous global symmetry having a Noether current, and that only non-compact generalized symmetries can be charged under a continuous global symmetry. The natural extension of these statements to finite group global symmetries is also described.

The discussion will make it clear that the weak and strong versions of Noether's theorem are not equivalent. There is a non-empty space of QFT's satisfying the weak version and violating the strong one. Further, we arrive at a potentially complete characterization of such space. Very explicitly, this is the space of QFT's with generalized symmetries charged under the global continuous symmetry.  These theories are very special though, having non compact generalized symmetries, what is in agreement with the almost universal validity of the strong version.

 These results are then used to provide a new proof of the Weinberg-Witten theorem. In the present light, the Weinberg-Witten theorem arises as a topological obstruction  for a global symmetry to have a conserved current because of the existence of one-form generalized symmetries charged under it. In this direction, the present framework allows generalizations of different sorts, such as different than four spacetime dimensions, massless particles in different representations of the little group, non-relativistic examples and extensions to finite-group global symmetries.

To illustrate the physics we explicitly study spin zero, one and two examples, with continuous and finite global symmetries. All the aspects described in the abstract general discussion, particularly those related to the new classification of twists operators, will appear in these concrete examples.

We end with a discussion of further expectations which we were not able to prove at this stage. This includes an avenue towards the proof that the complete characterization of QFT's violating the strong version of Noether's theorem is the one mentioned above, comments on the constrained dynamical nature of QFT's with non-compact generalized symmetries and intriguing connections with the Coleman-Mandula theorem.

The organization of the paper is as follows. In section~(\ref{dos}) we describe the local QFT, or algebraic approach to generalized symmetries that forms the basis for further developments. In section~(\ref{tres}) we develop the finer classification of twist operators arising for QFT's with generalized symmetries. We then use this classification to derive our main results. Section~(\ref{cuatro}) is devoted to a deeper technical analysis of twist operators in QFT's with generalized symmetries. It can be omitted on a first reading. In section~(\ref{cinco}) we illustrate the new classification of twist operators and the derived results with explicit examples, clarifying the nature of the violation of the strong version of Noether's theorem. In section~(\ref{ww}) we provide a different proof of Weinberg-Witten's theorem, and discuss several generalizations.  Section~(\ref{siete}) is devoted to a final discussion of several ideas/conjectures/applications in different areas that intersect with the present article.

\section{Local QFT approach to generalized symmetries}
\label{dos}

This section briefly reviews the local QFT approach to generalized symmetries developed in \cite{Casini:2020rgj,Review}. In particular we closely follow sections two and three of \cite{Review}. 

In the local algebraic approach to QFT \cite{Haag:1963dh}, the basic structure is the causal assignation of algebras of operators ${\cal A}(R)$ to spacetime regions $R$.  This is called a net of algebras. We will only need to talk about regions that, even if they have spacetime volume, are based on a single spatial plane $x^0=0$. Hence with $R$ we will mean a spatial region which has attached both a spacetime region (such as its causal development) and an algebra.

A natural assumption/starting point in this approach is that the operator content of the QFT is ultimately generated by local degrees of freedom, or field operators, localized in arbitrarily small regions of space. This property can be called ``additivity'' and we will assume it in what follows.\footnote{For QFT's defined in topologically non-trivial manifolds one typically finds operators that are not additive. These operators give rise to different superselection sectors of the algebra of local operators. In a given superselection sector, the local algebras are again additive.} This additivity principle appears transparently in lattice formulations of QFT in flat space. For example, Wilson or t' Hooft loops, often regarded as genuine line operators, can otherwise be constructed in an additive manner in balls containing the loops, even in non-abelian gauge theories. This is explained in app B of Ref. \cite{Casini:2020rgj}. In CFT's, as defined by the bootstrap approach, the full algebra of the theory is generated by the spectrum of local operators.

More generally, without assuming any particular QFT model, we can provide a mathematical form for this additivity principle by assuming that the algebra of a ball coincides with the algebra generated by any collections of smaller balls covering it. This allows the construction of an ``additive net'' by assigning to any region $R$ (with any given topology) the following additive algebra
\be
{\cal A}(R)\equiv \bigvee_{B \, \textrm{ball}\,, \cup B=R} {\cal A}(B)\,,\label{sdd} 
\ee  
where ${\cal A}_1\vee {\cal A}_2=({\cal A}_1\cup {\cal A}_2)''$ stands for the  von Neumann algebra generated by the two.  Here  ${\cal A}'$ is the algebra of all operators that commute with those of ${\cal A}$ (the commutant of ${\cal A}$) and the double commutant ${\cal A}''={\cal A}$ coincides with the algebra for any von Neumann algebra. Eq. (\ref{sdd}) is a self-consistent additive relation. The additive net satisfies, by construction,  the additive property
\be
{\cal A}(R_1\cup R_2)={\cal A}(R_1)\vee {\cal A}(R_2)\,,\label{3}
\ee 
and isotonia
\be
{\cal A}(R_1)\subseteq  {\cal A}(R_2)\,,\hspace{1cm}R_1\subseteq R_2\,. \label{isotonia}
\ee
The causality constraint on the additive net, embodying the commutativity of operators at spacial distance, takes the following form
\be
{\cal A}(R) \subseteq  ({\cal A}(R'))'\,,\label{causality}
\ee
where $R'$ is the causal complement of $R$, i.e. the set of points spatially separated from $R$.

Looking at the causality constraint~(\ref{causality}), we may ask whether the inclusion of algebras is saturated or not. If it is saturated we say ``duality'' holds for the region $R$ while if it is not we say duality is violated for such a region. If duality is violated, there must be at least two natural algebras associated with the same region $R$, namely ${\cal A}(R)$ and ${\cal A}(R')'$. The second one is the largest possible algebra associated with $R$ which still commutes with the observables in $R'$.  We will thus rename it as
\be
{\cal A}_{\textrm{max}}(R)\equiv({\cal A}(R'))' \;. 
\ee
The violation of the duality property of the additive algebra for a certain region $R$ is directly related to the appearance of generalized symmetries. From a generic perspective, to see this notice that such violation implies there must be a set $\{a\}$ of ``non-locally generated operators'' in the region $R$ such that
\be 
{\cal A}_{\textrm{max}}(R)={\cal A}(R)\vee \{a\}\;.
\ee
Below we call the operators $a$ the ``non-local operators'' in $R$ since they cannot be generated in a local manner inside $R$. Notice that the additivity principle implies that although one operator can be non-local in a certain region $R$, the same operator usually becomes locally generated in any ball containing $R$. 

The non-local operators $a$ may now be used to define irreducible classes/sectors $[a]$ of operators in ${\cal A}_{\textrm{max}}(R)$. More concretely, we define the class $[a]$ of ${\cal A}_{\textrm{max}}(R)$ to be the set of operators of the form $\sum_\lambda O^\lambda \,a\, \tilde{O}^\lambda$, with $O^\lambda$ and $\tilde{O}^\lambda$ belonging to ${\cal A}(R)$. The sector is said to be irreducible if there are no non-trivial subspaces inside it invariant under the simultaneous left and right action of the additive algebra. In the opposite case, it is said to be reducible and it can be decomposed into a sum of irreducible ones. Because ${\cal A}_{\textrm{max}}(R)$ is an algebra, the set of classes must close a fusion algebra between themselves $[a][a']=\sum_{a''} n^{a''}_{a a'} [a'']$. The identity class $[1]$ is by definition ${\cal A}(R)$, and its fusion rules are by construction $[1][a]=[a][1]=[a]$. These fusion rules might be related to the fusion of representations or conjugacy classes of a certain group, but they can be more general, the paradigmatic example being QFT's in two dimensions. 

It is an underlying natural assumption in QFT that ${\cal A}(R)$ has no center.\footnote{The center of an algebra is the set of operators commuting with all the operators in the algebra.} These algebras are called factors and are themselves irreducibles. We can go from any non zero element to any other by combinations of left and right actions with other operators belonging to ${\cal A}(R)$. In the same way $ {\cal A}_{\rm max}(R)$ has no center. It is irreducible with respect to arbitrary products of operators in $ {\cal A}_{\rm max}(R)$.  These conditions arise from the generic impossibility of smearing field operators on a $d-2$ dimensional boundary of the region (to act as a center element) and still producing a well defined operator in the Hilbert space. Both facts imply that we can obtain operators in ${\cal A}(R)$ from non local elements in $ {\cal A}_{\rm max}(R)$ by multiplying with other non local elements in $ {\cal A}_{\rm max}(R)$ and arbitrary left-right actions of operators in ${\cal A}(R)$.

Now, one of the key results of \cite{Casini:2020rgj,Review} is that the violation of duality in region $R$ forces a violation of duality in the complementary region $R'$. This follows from von Neumann's double commutant theorem  ${\cal A}''={\cal A}$. We conclude that if there are non local operators $\{a\}$ for $R$, there must be non local operators $\{b\}$ for the complement $R'$, 
 \be 
{\cal A}_{\textrm{max}}(R')={\cal A}(R')\vee \{b\}\;.
\ee
These non local operators must have their own fusion rules $[b][b']=\sum_{b''} n^{b''}_{b b'} [b'']$. Because  ${\cal A}_{\textrm{max}}(R)={\cal A}(R')'$ the non local operators of $R$ commute with the additive operators in $R'$ and viceversa. However, the non local operators $\{a\}$ and $\{b\}$ for $R$ and $R'$ cannot commute (all of them) with each other. Such commutation would imply ${\cal A}_{\textrm{max}}(R)\subseteq ({\cal A}_{\textrm{max}}(R'))'={\cal A}(R)$, which is not possible if the inclusions ${\cal A}\subset {\cal A}_{\textrm{max}}$ are strict. This failure of commutativity is not a failure of causality since those operators cannot be constructed locally in their respective regions.

It is a natural assumption in QFT that the generalized symmetries are transportable. This means that smoothly deforming a region $R_1$ to another one $R_2$ with the same topology, such that the tube of homotopy  $T$ described by the deformation has also the same topology, the non local operators in  $R_1$ can be converted to non local operators in $R_2$ by additive operators in $T$. This notion of generalized symmetries is topological in this sense, with the non local classes preserved under deformations. We will assume transportability in what follows.\footnote{ For generalized symmetries generated by fluxes of physical $k$-forms this transportability is automatic. Though this issue has not been analyzed in the literature, it is quite plausible that transportability follows from the existence of additive twists for translations, see below.}

In \cite{Magan:2020ake,Casini:2020rgj}, the previous algebraic structure specifying the violation of duality for regions $R$ and $R'$ was summarized by a quantum complementarity diagram
\bea\label{cdiaor}
{\cal A}_{\textrm{max}}(R)\equiv {\cal A}(R)\vee \{a\} & \supset &{\cal A}(R)\nonumber \\
\updownarrow\prime \!\! &  & \,\updownarrow\prime\\
{\cal A}(R')& \subset & {\cal A}(R')\vee \{b\}\equiv {\cal A}_{\textrm{max}}(R')\,.\nonumber 
\eea
In this diagram, on the upper side, we have the inclusion of the two algebras associated with region $R$. Going down in the diagram amounts to taking commutants, producing the algebras naturally associated with the complementary region $R'$. In QFT we can assume that for these type of inclusions ${\cal A}(R)\subset {\cal A}_{\rm max}(R)$  the relative commutant is $ {\cal A}_{\rm max}(R)\cap {\cal A}(R)'= {\cal A}_{\rm max}(R)\cap{\cal A}_{\rm max}(R')=\{1\}$, namely the identity algebra. This is equivalent to strong additivity, $({\cal A}(R)\vee {\cal A}(R'))'=\{1\}$. Physically, this is because for a non local operator to commute with all the additive algebra it has to be sharp, namely, without smooth smearing in the region. This again makes the operator too singular to be well defined. Pathological scenarios such as generalized free fields could defy this feature but we will not consider those here.

The non-local operators for each region (the $a$'s and $b$'s) generate transformations of the maximal algebras of the complementary regions. In this sense, the non-local operators act as generalized symmetry operations, where the charged objects are the complementary non-local operators and the neutral objects are the observables in the additive algebra.

The relation with the generalized global symmetries defined in \cite{Gaiotto:2014kfa} was described in \cite{Casini:2020rgj,Review}. The main difference, that might cause confusion, is that Ref \cite{Gaiotto:2014kfa} defines the generalized symmetry to be related to the choice of a maximum set of commuting non local operators for any region. In Ref \cite{Review} this was called a Haag-Dirac net of algebras.  For the case of gauge theories, this choice is said to define the global  gauge symmetry group. From the local QFT perspective, all the choices of non local operators are physically equivalent and it proves more suitable in certain circumstances to work without making such a specific choice.  In fact, the non local operators cannot be eliminated from the theory but only from the operator content of a specific region, since they are generated locally in topologically trivial regions. So in the present context, the generalized symmetries actually contain all the possible non-local operators.   This is just a terminological difference with \cite{Gaiotto:2014kfa} that does not entail any difference in the involved physics.

This local approach to generalized symmetries in QFT, based on the violation of duality of the additive algebra, might appear somewhat abstract on a first encounter. But it has by now been studied in several explicit examples, both old and new, with both discrete and continuous groups and for different $p$-form symmetries. For an extensive treatment of QFT's with global symmetries see \cite{Casini:2019kex,Magan:2021myk}, where the appareance of generalized symmetries was shown to reflect on new universal terms in the entanglement entropy. For non-abelian gauge theories, both in the continuum and in the lattice see \cite{Casini:2020rgj,Pedro}. For applications to aspects of quantum gravity see \cite{Review}. For the graviton field see \cite{casini2021generalized}, which contains some key motivations for the present work, which will be described in a later section. At any rate, below we will consider new explicit examples.

\section{Global symmetries, Noether currents, and generalized symmetries}
\label{tres}

We now introduce a new ingredient to the previous approach to generalized symmetries. We start with a QFT with certain additive algebra $\mathcal{A}(R)$, possibly violating duality in certain regions, and therefore with potential generalized symmetries. But we now let this additive algebra to be charged under a global symmetry group $G$.\footnote{Related scenarios with mixing of symmetries have appeared recently in \cite{MixingSym1,MixingSym2} and references therein.} 

To start with consider $G$ to be an internal global symmetry. By definition, the group acts as automorphisms of the additive algebras for any $R$. For an unbroken global symmetry these are implemented by unitaries 
\be\label{invalg}
U(g)\, {\cal A}(R)\, U(g)^{-1}={\cal A}(R)\,,\hspace{.7cm} g\in G\,.
\ee 
This relation, and the ones that follow, is to be understood as a mapping between algebras. It is not saying that all elements of ${\cal A}(R)$ are invariant under the symmetry group, but that they transform between themselves. It is easy to verify that the conjugation with the unitary $U(g)$ carries commutant algebras into commutant algebras, namely, given~(\ref{invalg}) we have that
\be\label{invalga}
U(g)\, {\cal A}(R)'\, U(g)^{-1}={\cal A}(R)'\,,\hspace{.7cm} g\in G\,.
\ee 
Since ${\cal A}(R')'={\cal A}_{\rm max}(R)$, this conjugation is also an automorphism of the maximal algebras
\be
U(g)\, {\cal A}_{\rm max}(R)\, U(g)^{-1} ={\cal A}_{\rm max}(R)\,.
\ee
Therefore, the symmetry cannot convert non local operators into local ones or viceversa. It has to transform the non local operators into themselves. The question we want to study is whether this group action can change the non local classes of a given region $R$ or must leave the classes invariant. And then seek for the consecuences in each scenario.

\subsection{Point-like transformations of class labels}\label{pointl}

We start by considering the possible transformations of class labels. We choose representatives for the non local classes $[a_\lambda]$ in $R$, say $a_\lambda$. For the identity class we can choose the identity operator itself. We remind the classes $\left[ a_\lambda \right]$ are disjoint sets of operators invariant under the left and right action of  ${\cal A}(R)$, and they are irreducible when there are no proper subsets of $\left[ a_\lambda \right]$ invariant under ${\cal A}(R)$. Therefore, by definition, any operator $A\in {\cal A}_{\rm max}(R)$ can be written as 
\be
A=\sum_{\lambda,s}\, O_{\lambda,s}\, a_\lambda\, \tilde{O}_{\lambda,s}\,,
\label{cc}
\ee
where $O_{\lambda,s}, \tilde{O}_{\lambda,s} \in {\cal A}(R)$.

For a non irreducible combination of classes such as~(\ref{cc}),  we can project the sum to one of the non local irreducible operators appearing with non trivial coefficient on the right hand side of (\ref{cc}). We do this by acting on the left and on the right with combinations of elements of  ${\cal A}(R)$ (using projectors constructed from dual non local operators \cite{Casini:2020rgj} that fit into $R$)
\be\label{proj}
\sum_i \,P_i \,A\, \tilde{P}_i= a_{\lambda}\,,\hspace{.7cm} P_i,\tilde{P}_i \in {\cal A}(R)\,.
\ee
Now we consider again the global symmetry. The transformation of a non local operator $a_\beta$ under the global symmetry belongs to ${\cal A}_{\rm max}(R)$. Therefore it can be written as 
\be
U(g)\, a_\beta \, U(g)^{-1}= \sum_\lambda O_{\lambda,s}\, a_\lambda \, \tilde{O}_{\lambda,s}\,. \label{dec}
\ee
As before, we can project the right hand side into one $a_\lambda$ by multiplying this equation on the left and on the right by certain $P_i$ and $\tilde{P}_i$ belonging to $ {\cal A}(R)$ and sum over $i$. We obtain
\be
U(g)\, \left(\sum_i Q_i \, a_\beta\, \tilde{Q}_i\right) U(g)^{-1}= a_\lambda\,,\label{solo}
\ee
where
\be
 Q_i =U(g)^{-1}\, P_i \, U(g)\in {\cal A}(R)\,, \,\,\,\,\,\,\,\,\,\tilde{Q}_i =U(g)^{-1}\, \tilde{P}_i \, U(g)\in {\cal A}(R)\,,
\ee
so that 
\be 
\left(\sum_i Q_i \, a_\beta\, \tilde{Q}_i\right) \in [a_\beta]\;.
\ee
Therefore, eq.~(\ref{solo}) says that an element of the class $\beta$ is transformed into an element of the class $\lambda$ alone. This, together with the fact that the global symmetry maps $ {\cal A}(R)$ into $ {\cal A}(R)$, it follows that all the elements belonging to class $\beta$ are transformed into the elements belonging to class $\lambda$. Hence, there can only be one non zero class in the decomposition (\ref{dec}). We conclude that the group acts as a point transformation in the manifold of class labels, which is divided into orbits under the action of the global symmetry group, leaving the identity class invariant. For finite groups these orbits give permutation representations of the group. These point transformations must be a symmetry of the fusion rules associated with the classes $[a_\lambda]$.  It also follows that the action of the group on non local classes is transportable (i.e. it is kept the same) under the identification of irreducible classes when we continuously  deform the region but not its topology.   

If the global symmetry is spontaneously broken (SSB) the group automorphisms cannot be implemented by a global unitary operator $U(g)$, but inside any fixed bounded region $R$ it can be implemented by a unitary \cite{buchholz1992new}. Local unitaries effecting the group transformation in compact regions $R$ are the twists $\tau_g (R)$. These twists will play a central role in what follows and we will be discussing them next. Using these twists, it is not difficult to show the conclusions of this section still hold for the case of spontaneously broken global symmetries since the generalized symmetries can always be studied inside a finite region $R$ with the topology of a ball.

\subsection{Twist and their different types}
\label{types}

\begin{figure}[t]
\includegraphics[width=.45\linewidth]{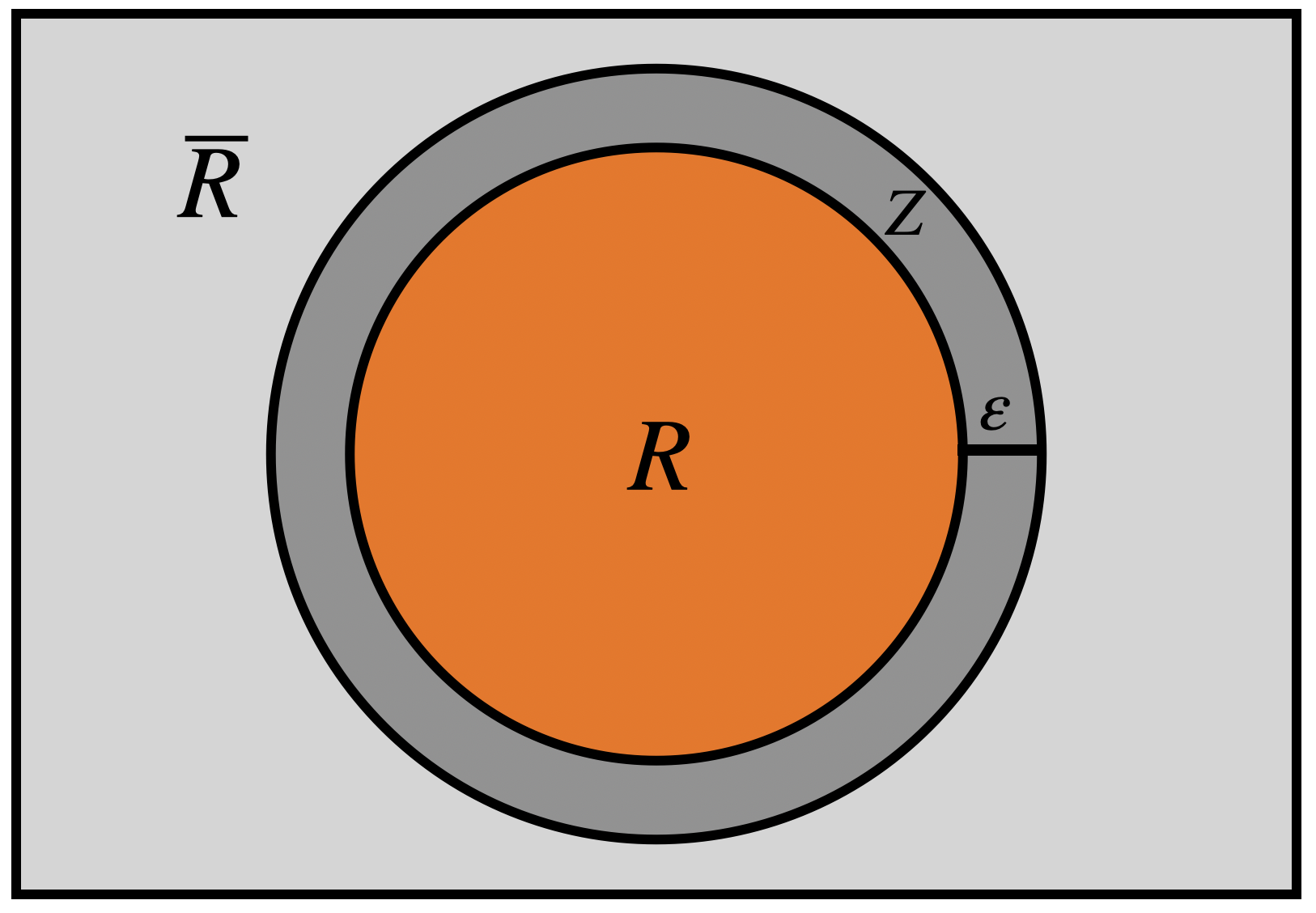}
\centering
\caption{Geometric configuration characterizing the definition of a twist operator $\tau_g(R,Z)$. The topology of the region $R$ is not required to be trivial for the twist to be defined. The twist is constructed so as to effect the symmetry transformation in ${\cal A}(R)$ and leave the operators in ${\cal A}(\bar{R})$ invariant. The buffer zone $Z$ with non-zero size $\epsilon$ is required for the operator to exist in the QFT.}
\label{Figtwist}
\end{figure}

To proceed we now introduce the twist operators. These are  local unitaries effecting the group transformation only in a certain region $R$. We now discuss them in more detail. Given a region $R$ we can define another slightly bigger region $R\cup Z$,  such that its boundary is separated from the one of $R$ by a sufficiently small distance $\epsilon$ ($\epsilon$ is the width of $Z$). The region $Z$ will be called the buffer or the smearing zone. The boundary of $Z$ is $\partial Z=(\partial Z)_1\cup (\partial Z)_2$, with $(\partial Z)_1=\partial R $ and $(\partial Z)_2=\partial (R\cup Z)$. The topologies of $(\partial Z)_1$ and $(\partial Z)_2$ are the same.  The topology of $Z$ is  the same as $\partial R\times \mathbb{R}$. We also call $\bar{R}=(R\cup Z)'$, such that $R\cup Z\cup \bar{R}$ is a partition of the space by three disjoint regions. This geometric configuration is depicted in Fig~(\ref{Figtwist}).

With these geometric conventions, a twist $\tau_g(R,Z)$ is a unitary operator that implements the global symmetry operation $g$ (by conjugation) on the operators of the algebra ${\cal A}(R)$ and leaves the operators in ${\cal A}(\bar{R})$ invariant.  The buffer zone was introduced by the technical reason that as $\epsilon$ collapses to zero the twist is too singular to be an operator. By definition, from its trivial action on ${\cal A}(\bar{R})$, it follows that
\be\label{taugmax}
\tau_g(R,Z)\in {\cal A}_{\rm max}(R\cup Z)\;.
\ee
The intuition associated with these operators gets clarified whenever we have a Noether current $j^\mu$. In these cases we integrate the charge over $R$ with smearing equal to one, and then let the smearing die in the buffer zone $Z$. But, as described in the next section, these twists exist under very general conditions, even for finite groups, and can be constructed using modular theory.

When the  QFT in which the symmetry group $G$ acts displays generalized symmetries, we might have strict inclusions of the sort discussed in the previous section, namely $ {\cal A}\subset  {\cal A}_{\rm max}$. In particular, for the defining region of the twist $R\cup Z$ we might have 
\be
{\cal A}(R\cup Z)\subset  {\cal A}_{\rm max}(R\cup Z)\;,
\ee
due to the existence of non-local operators in $R\cup Z$. 

In these scenarios, there are refinements on the possible species of twists we can have. First, a twist can belong to the additive algebra ${\cal A}(R\cup Z)$, instead of just belonging to the maximal one~(\ref{taugmax}). In this case we say the twist is ``additive''. If it is not additive, it is by definition a non local operator of ${\cal A}_{\rm max}(R\cup Z)$. Second,  a twist can implement the global symmetry on the maximal algebra ${\cal A}_{\rm max}(R)$, instead of just the additive one. In this case we say the twist is ``complete''. Notice that a general twist is not required to act on the non-local operators in $R$ since these operators only become local in a region that goes beyond the one defining the twist. Below we will discuss examples of additive/non-additive and complete/non-complete twists. Further refinements will be discussed in the next section, related to the way the twist transforms the different algebras in the buffer zone. But the characterization in terms of complete and additive twists will be enough for understanding most of the results of the article.

Given a twist $\tau_g(R,Z)$ for $R$, we can construct a ``complementary'' twist $\tau_g(\bar{R},Z)$ for $\bar{R}$ as
\be
\tau_g(\bar{R},Z)=U(g)\, \tau_g(R,Z)^{-1}\,,
\ee
where $U(g)$ is the global representation of the symmetry group. It follows from the definition that $\tau_g(\bar{R},Z)$ indeed implements the group operations on ${\cal A}(\bar{R})$, leaving the operators in ${\cal A}(R)$ invariant. If $\tau_g(R,Z)$ is complete, the complementary twist $\tau_g(\bar{R},Z)$ also leaves non local operators in ${\cal A}_{\rm max}(R)$ invariant since every non-trivial transformation is canceled by $U(g)$. Hence, it belongs to the commutant of ${\cal A}_{\rm max}(R)$, namely ${\cal A}(\bar{R}\cup Z)$. Therefore,  $\tau_g(\bar{R},Z)$ is an additive twist for $\bar{R}$ if $\tau_g(R,Z)$ is complete. In the same way, if $\tau_g(R,Z)$ is additive, the complementary twist is complete. Hence, complete and additive twists are complementary or dual notions.  

It is immediate that the existence of twists $\tau_g(R,Z)$ for $R$, that are simultaneously complete and additive, implies the non local classes corresponding to $R$ are invariant under $G$. The generalized symmetry is therefore uncharged with respect to $G$. This follows from the following chain of arguments. The twist being complete, it implements the group operation on non local operators in $R$. Non local operators in $R$ are non local operators in $R\cup Z$ by transportability. The twist being an additive operator in $R\cup Z$, it cannot effect transitions between the classes of $R\cup Z$. Therefore the generalized symmetry is not charged under $G$. Of course, the complete and additive twist can change the additive content of the non local operators since additive operators are generically charged under $G$. In section \ref{cuatro} we show the converse statement: uncharged classes implies the existence of complete additive twists (assuming the split property). As we will discussed in serveral places through the article, both statements are crucial to undertand the difference between the weak and strong versions of Noether's theorem.

In consequence, if there are non local classes in $R$ non trivially transforming (charged) under $G$ we have two different possibilities. Either the twist is non-complete or, if it is complete, it must be non additive. In the second case the twist has to be a non local operator in $R\cup Z$ itself, namely the twist contains non local elements of ${\cal A}_{\rm max}(R\cup Z)$. As in general non local operators in $R$ can be transported to the buffer zone $Z$, we can as well think the twist belongs to ${\cal A}_{\rm add}(R)\vee {\cal A}_{\max}(Z)$. In this presentation the twist appears to have ``non local boundary terms''.\footnote{We remark this should not be confused with a UV regularization problem. The fact that the correction is non-local is a macroscopic feature that does not depend on UV ambiguities and that it has consequences for IR physics.}

\subsection{The case of continuous spacetime symmetries}

Spacetime symmetries $\Lambda$ introduce the novelty that the algebras are not kept invariant but are transformed geometrically
\be
U(\Lambda)\,  {\cal A}(R)\, U(\Lambda)^{-1}={\cal A}(\Lambda\,R)\,. 
\ee
For continuous spacetime symmetries and small enough transformations the regions get a little displaced or distorted but, because of transportability, there is a unique identification of non local sectors between the transformed algebras. Therefore, the question of the change or not of the non local sectors under the symmetry operation makes perfect sense: it is the question of whether the unitary transformation implementing the symmetry changes classes with respect to the ones associated by transportability. 

The definition of twists can be generalized for finite spacetime symmetry transformations by requiring that the twist implements the symmetry operation on ${\cal A}(R)$ only on operators $O \in {\cal A}(R) $ such that $U(\Lambda)\, O \,U(\Lambda)^{-1}$ is still in ${\cal A}(R)$. Standard twists can be constructed for this case as well using the split property \cite{buchholz1986noether}, as we review in detail in the next section. For our purposes, it will be enough to talk about the local charges,  infinitesimal generators of the twists for continuous symmetries.  These charges are required to have the same commutators as the global charge with elements in ${\cal A}(R)$  and commute with elements in ${\cal A}(\bar{R})$. Additive charges are affiliated\footnote{Since a charge is in general not a bounded operator it cannot belong to the von Neumann algebra, but instead it is affiliated to it if the spectral projectors belong to the algebra.} to the additive algebra of $R\cup Z$, namely ${\cal A}(R\cup Z)$, and complete ones have the same commutator as the global charge with the maximal algebra of $R$, namely ${\cal A}_{\rm max}(R)$.  

\subsection{Generalized symmetries cannot be charged under Noether charges}
\label{noetherz}

Previously we showed, from a general standpoint, that the existence of simultaneously additive and complete twists imply the generalized symmetry is uncharged under the global symmetry group. Now we will show that global symmetries implemented by a Noether current always have additive and complete twists. Therefore,  global symmetries implemented by a Noether current must leave invariant the non local classes of all generalized symmetries. Equivalently, generalized symmetries must be uncharged under global symmetries implemented by Noether currents. Conversely, this demonstrates an obstruction to the existence of Noether currents when there are non local classes transforming under a continuous symmetry. It shows that the space of QFT's violating the strong version of Noether's theorem includes all the QFT's with generalized symmetries charged under a global continuous symmetry group.\footnote{We will later argue that that this is indeed the complete characterization of QFT's violating the strong version of Noether's theorem.}

\begin{figure}[t]
\includegraphics[width=0.7\linewidth]{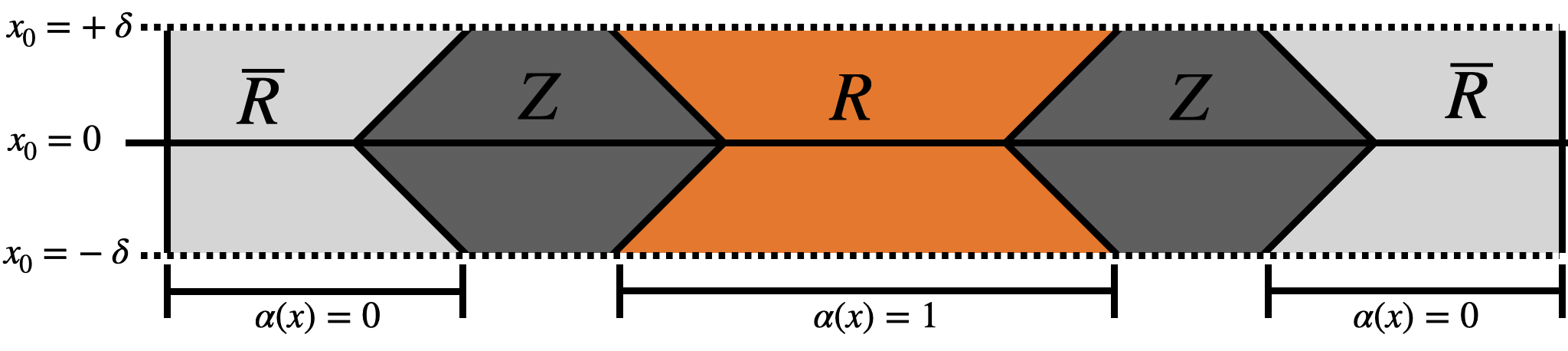}
\centering
\caption{Geometric configuration used for the definition of twist operators out of Noether currents. The smearing function $\alpha(\vec{x})$ is equal to one for the future and past of $R$ in the time slice $(-\delta,\delta)$ and is equal to zero in the future ans past of $\bar{R}$ in the same time slice.}
\label{rt}
\end{figure}

Consider a region $R$ (with any given topology) with buffer zone $Z$, and assume the global symmetry is generated by a local Noether current $j_{\mu}$. A twist $\tau_g (R,Z)$, implementing the symmetry in $R$ and doing nothing in $(R\cup Z)'$, can be constructed by smearing the Noether current $j_{\mu}$. Explicitly, we can define the local twist and charge as
\be
\tau_\lambda (R,Z)= e^{i\, \lambda\, Q(R,Z)}\,, \hspace{.6cm} Q(R,Z)=\int d^{d}x\, \beta(x^0)\,\alpha(\vec{x})\,j_0(x)\,. \label{rty}
\ee
In this formula $\lambda$ determines the specific element of the symmetry group $G$. The smearing functions $\alpha$ and $\beta$ are smooth. They further satisfy the following requirements: $\int dx^0\, \beta(x^0)=1$,  the support of $\beta$ is in $[-\delta,\delta]$ for $\delta$ small enough, $\alpha(\vec{x})=1$ for all spatial coordinates of the points inside the intersection of the future and past light cone of $R$ with $x^0\in [-\delta,\delta]$, $\alpha(\vec{x})=0$ for all points inside the intersection of the future and past light cone of $\bar{R}=(R\cup Z)'$ with  $x^0\in [-\delta,\delta]$.  This geometric configuration is depicted in Fig~(\ref{rt}). This choice of smearing functions ensures that the commutator of the localized charge $Q(R,Z)$ with any local operator in $R$, and therefore with the additive algebra $\mathcal{A}(R)$ of region $R$, coincides with the commutator of the global charge operator. It also ensures that the commutator of $Q(R,Z)$ with local operators in region $\bar{R}$ vanishes. The operator $\tau_\lambda (R,Z)$ defined above is then a twist operator for region $R$ and buffer zone $Z$.

It is evident that Noether twists $\tau_g(R,Z)$, being formed locally by the algebra of a local field operator, are always additive in $R\cup Z$. It is now simple to show that Noether twists are also necessarily complete. Let $Q$ be the global charge, having  the same expression~(\ref{rty}) but where $\alpha(\vec{x})=1$ for all $\vec{x}$. This charge $Q$ generates the global symmetry transformation for all non local operators in $R$ since these operators are ultimately additive in the full theory. We have trivially $Q=Q(R,Z)+(Q-Q(R,Z))$. From the definition of the localized charge~(\ref{rty}), the operator $Q-Q(R,Z)$ is additive in $\bar{R}\cup Z$. It therefore commutes with all non local operators in $R$. This implies that the commutator of $Q(R,Z)$ with the non local operators in $R$ is the same as the one of $Q$. Therefore the Noether twists are additive and complete. We conclude that for QFT's in which the symmetry is generated by a Noether current, the non local classes, associated with any region of general topology, are invariant under $G$.

Continuous spacetime symmetries (Poincar\'e and conformal symmetries) are generated by the stress tensor when this later exists. We can generate a Poincar\'e or conformal twist charge using the stress tensor just as described above. 
 Therefore, if the QFT contains a stress tensor, the corresponding local charges will be additive and complete. We conclude that for theories with a stress tensor, the generalized symmetry classes cannot be charged under continuous spacetime symmetries.\footnote{Related arguments have been used in \cite{buchholz1986noether,Longo:1994xe} to show that the global symmetry group underlying a structure of charge superselection sectors commutes with Poincar\'e transformations when there is a stress tensor. Those are a particular case of the present theorem when restricted to the case of two ball sectors, or orbifold sectors. See the examples in section \ref{orbi}. Here the result is shown to extend to all generalized symmetries.}

\subsection{Non compact generalized symmetries}
 
If generalized symmetry classes cannot be charged under continuous symmetries implemented by a Noether current, it is then interesting to investigate the consequences that may carry the existence of generalized symmetries that are charged under a continuous symmetry. 

Given the point-like transformations between classes described earlier, the set $H$ of elements of $G$ leaving invariant all non local classes is a normal subgroup of $G$.  We can focus on the quotient $\tilde{G}=G/H$ then.  Except the identity, none of its elements leave all the classes invariant. We are concerned with the case where $\tilde{G}$ is a Lie group. This group will act pointwise in the set of class labels. Therefore, the non local classes for $R$ must form a continuum.  Under the action of $\tilde{G}$ such manifold might break into different orbits, generated by the group and a single point in the orbit. If such point has a non-trivial stabilizer group, the dimension of the orbit is the dimension of the quotient of $\tilde{G}$ by the stabilizer group. There must be at least a one dimensional orbit or the classes would be invariant. 

Below we show that If the non local classes of $R$ are not invariant, the non local classes for $R'$ must necessarily be non invariant too.  Then they must also form a continuum of classes. We conclude that when generalized symmetries are charged under a continuous global symmetry group, the generalized symmetries must have a continuum of non commuting dual classes. 

We remark this is a strong constraint for these QFT's. These scenarios can be made into a definition of what is meant by a non compact generalized symmetry.\footnote{Notice for a compact group the conjugacy classes form a continuum but the representations are discrete.} In the examples we discuss below, the models having non compact generalized symmetries are very special: they are massless and free. Hence, the models we find having a continuous symmetry and no Noether current are also very special.  Though we could not prove it, we suspect this should be a general feature of non compact generalized symmetries, disregarding the existence of any global symmetry. We will say a bit more about this problem in the discussion at the end of the paper. 

\section{Further properties of twists and standard twists}
\label{cuatro}

The previous section contained the two main important results: generalized symmetries charged under a continuous global symmetry are not compact, and in such scenarios the global symmetry group cannot be generated by a Noether current. With this information, and the finer classification of twists in terms of additivity and completeness, the reader can safely jump to the next sections where we discuss explicit examples and generalizations of the Weinberg-Witten theorem. The present section is more technical than the previous one and it is not needed to understand the main points of the rest of the paper. It will be important though in relation to a potential complete classification of the QFT's violating the strong version of Noether's theorem. It also completes the analysis of the classification of twists operators when generalized symmetries are present.

We start by reviewing the construction of standard twists using the split property. This ensures the existence of twists in most theories of physical interest. We extend the construction to cases where generalized symmetries are present. We then study the non local sectors on the buffer zone and the action of the twists on this region. This allows us to find new sufficient conditions for having twists that are simultaneously additive and complete. Using this analysis we then derive the main result of this section. We prove that there are complete additive twists when the classes are uncharged and that they can be constructed in a standard manner. This is the converse of the statement derived previously. We end with a discussion on the possibility to concatenate small twists to produce others for larger regions.  

\subsection{Standard twists constructed from the split property}
\label{standard1}

There is a standard though abstract way to construct twist operators \cite{doplicher1982local,Doplicher:1983if,Doplicher:1984zz,buchholz1986noether}. It requires the split property. We review the original way in which this construction was formulated, and then introduce the variations that appear when there are generalized symmetries in QFT on top of the global symmetry. We will follow the construction on \cite{buchholz1986noether} that can be applied more generally.

Given two commuting algebras  ${\cal A}$ and ${\cal B}$,  and a state $\vert\Omega\rangle$ cyclic and separating for ${\cal A}\vee{\cal B}$, the split property asserts the existence of a type I factor ${\cal N}$ such that ${\cal A}\subset {\cal N}$, ${\cal B}\subset {\cal N}'$. A type I factor is the algebra of all bounded operators in some Hilbert space. An equivalent description of the split property is that the algebras ${\cal N}$ and ${\cal N}'$ are the algebras of the operators acting in each of the two factor Hilbert spaces $\mathcal{H}_{\cal N}$ and $ \mathcal{H}_{{\cal N}'}$ of a decomposition of the full Hilbert space as a tensor product
\be
\mathcal{H}=\mathcal{H}_{\cal N}\otimes \mathcal{H}_{{\cal N}'}\,.
\ee
The algebras ${\cal A}$ and ${\cal B}$ are then included in the operator algebras acting on each side of this tensor product.

The split property for the algebras of spatially separated regions is expected to hold with great generality in QFT. It follows from two premises \cite{buchholz1986causal,buchholz1987universal}.  The first one is that time translations of  one of the algebras for any time in some non empty interval $(-t_0,t_0)$ still commutes with the other algebra. In relativistic local QFT, this is warranted by the existence of the buffer zone $Z$ between the two spatially separated regions, see Fig~(\ref{Figtwist}). The second one is a UV condition implying that the number of degrees of freedom at high energies does not increases too rapidly. In more concrete physical terms, this requires that the local free energy increases at most with a power law in temperature.  This condition ensures normal thermodynamic properties at all temperatures (no Hagedorn maximal temperature for example). Therefore, in the context of this paper, we will assume the split property to hold.

The split factor ${\cal N}$ in the split property is highly non unique. However, there is a general construction by Doplicher and Longo for a standard split \cite{Doplicher:1984zz}. It starts with a state $|\Omega\rangle$ that is cyclic and separating for ${\cal A}\vee {\cal B}$, and the induced modular conjugation $J\equiv J_{{\cal A}\vee{\cal B}}$ with respect to such algebra. In QFT it is natural to use the vacuum state. The standard split factor reads explicitly
\be
{\cal N}={\cal A}\vee (J \,{\cal A}\,J)\,, \hspace{.7cm} {\cal N}'={\cal B}\vee ( J \,{\cal B}\,J)\,.
\ee 
A nice feature of this construction is that ${\cal N}$ depends only on the vacuum and the algebras ${\cal A}$ and ${\cal B}$. 

Once we have a split, it is possible to choose a tensor product vector $|\eta\rangle=|\Omega\rangle_{\cal N}\otimes |\Omega\rangle_{{\cal N}'}$ such that the state induced by $|\eta\rangle$ on ${\cal A}$ and  ${\cal B}$ coincides with the product state induced by $|\Omega\rangle$ on these algebras. In other words, $|\eta\rangle$ is a purification of the vacuum state in each algebra inside the corresponding type I factors.  Notice this state does not have correlations between the algebras, in contrast to $|\Omega\rangle$. Further, the vector $|\eta\rangle$ is unique if chosen in the standard cone ${\cal P}$. The standard cone is defined as the set of vectors\footnote{See \cite{haag2012local} for a detailed exposition.} 
\be
 O\, J\, O \,|\Omega\rangle\,,\hspace{1cm} O\in {\cal A}\vee {\cal B}.    
\ee
More precisely, there is an invertible isometry $W: {\cal H}\rightarrow {\cal H}\otimes {\cal H}$ such that
\bea
W\, A\, B\,|\eta\rangle &=&  A\,|\Omega\rangle\otimes B\, |\Omega\rangle\,,\hspace{.6cm}A\in {\cal A}\,, B\in {\cal B}\,,\\
W \, A \, W^* &=& A\otimes 1\,,\hspace{.6cm}W \, B \, W^* = 1\otimes B\,,\hspace{.6cm}A\in {\cal A}\,, B\in {\cal B}\,.
\eea
This mapping carries the type I factors ${\cal N}$ and ${\cal N}'$ into the operator algebras of the two Hilbert spaces. It also follows $J\,|\eta\rangle=|\eta\rangle$, $W,J\, W^*= J_{\cal A}\otimes J_{\cal B}$.

This structure makes it transparent the construction of twists. We just have to use the previous construction with the associations
\be
{\cal A}\rightarrow {\cal A}(R)\,,\,\,\,\,\,\,\,\,\,\,\,\,\,\,\,\,\,\, {\cal B}\rightarrow {\cal A}(\bar{R})\;,
\ee
such that ${\cal A}(R)\subset {\cal N}$, ${\cal A}(\bar{R})\subset {\cal N}'$. 
For any unitary operator $U(g)$ acting in the global Hilbert space \cite{buchholz1986noether} defines 
\be
\tau_g(R,Z)=W^*\, (U(g) \otimes 1)\, W 
\ee
that acts non trivially only on the factor ${\cal N}$ and hence commutes with ${\cal A}(\bar{R})$. Further, if $U(g)$ carries an element $A$ of ${\cal A}(R)$ into another element $\tilde{A}$ of ${\cal A}(R)$ it is direct that $\tau_g(R,Z)$ will have the same action on $A$. Then,  $\tau_g(R,Z)$ gives a twist both for internal symmetries and spacetime symmetries. Further, from its definition, the twists $\tau_g(R,Z)$  furnish a representation of the group 
\be\label{stt}
 \tau_g \tau_{h}=\tau_{gh}\,.
\ee
This group product relations are not necessary for the twist to implement the group operations locally. Indeed, they typically do not hold for a more general  smearing of the twist in the buffer zone $Z$, as the ones arising by smearing Noether currents in conventional manners.  

 For an unbroken internal symmetry, the twists are just the restrictions of $U(g) $ to the two commuting type I factors, which are left invariant by the action of $U(g)$. We have   
 \be\label{splitsg}
U(g)=\tau_g^{\cal N}\otimes \tau_g^{{\cal N}'}= \tau_g(R,Z)\,\tau_g(\bar{R},Z)  \;.
\ee
Then these twists are complementary. It will be important later that the standard twists for internal symmetries also transform covariantly
\be\label{stt1}
 U(g) \,\tau_{h}(R,Z)\, U(g)^{-1}=\tau_{g h g^{-1}}(R,Z)\,,
\ee
a convenient feature that is particular of this construction.

In the case of spontaneously broken global symmetries we do not have a globally defined $U(g)$ implementing symmetry transformations. However, given any compact region $R$, it is always possible to construct a unitary operator $U_R(g)$ in the global Hilbert space that implements the symmetry in $R$, see \cite{buchholz1992new}. Using this unitary, the same construction above gives a twist $\tau_g(R,Z)$ for ${\cal A}(R)$ in this case. The lack of the global $U(g)$ does not interfere with our present investigation since generalized symmetries can be described within regions inside a ball and their relative complements inside the same ball.  A Hilbert space and type I factor for the ball  separated from the rest of the space by a split can be defined,  and one can restrict attention to the physics in this algebra. The only remaining difference with the unbroken symmetry case is that the vacuum state is non invariant under the group. Since we are not interested in the state but on the symmetries of the algebras, this can be dealt with by averaging the group transformations of the vacuum  state in a ball, and taking a purification inside in the split ball. 

Now we turn to the case when the QFT displays generalized symmetries for certain classes of regions $R$. In these QFT's we can consider more than one algebra for the same $R$, namely ${\cal A}(R)$ or ${\cal A}_{\rm max}(R)$.  Then, instead of starting the Buchholz-Doplicher-Longo construction with the commuting algebras ${\cal A}(R)$ and ${\cal A}(\bar{R})$, we can take the also commuting  algebras ${\cal A}(R)$ and ${\cal A}_{\rm max}(\bar{R})$. These should be split by the same reasons described above.  Therefore, the construction of the standard twists follows analogously. It is immediate that these twists are complete for $\bar{R}$ and hence additive for $R$. 
On the other hand, if we start with a standard split between ${\cal A}_{\rm max}(R)$ and  ${\cal A}(\bar{R})$, such standard twist will be complete for $R$ and additive for $\bar{R}$.\footnote{A natural expectation for the standard twists that arise from ${\cal A}(R)$ and ${\cal A}(\bar{R})$ is that they are in general not necessarily complete nor additive.}

This ends the review of the algebraic construction developed in \cite{doplicher1982local,Doplicher:1983if,Doplicher:1984zz,buchholz1986noether}, and its extension to include generalized symmetries. The outshot is that, given the split property in QFT, one can find the local twists $\tau_g (R,Z)$ for any region $R$, and smearing region $Z$. These are the local versions of the global symmetry operators $U(g)$. This proves the weak version of Noether's theorem. Moreover this construction applies as well to the case of discrete symmetries. However, as stressed in \cite{buchholz1986noether}, the derivation of a conserved current from the existence of twists is not warranted for continuous symmetries. Indeed, we will review a variety of simple examples that show this is not always possible, and that the strong version of Noether's theorem is not always valid.\footnote{Previously known examples were described in \cite{Current1,Harlow:2018tng}. The graviton is another prominent example as we discuss below. All of them fall into the class unraveled in this article, namely those QFT's with generalized symmetries charged under continuous symmetries.} More precisely, as discussed above, when generalized symmetries are present, there is an important difference in the nature of the standard twists from the ones arising through a putative Noether current. In particular, while standard twists can generally be constructed, complete additive twists cannot be constructed when the generalized symmetry is charged under the global symmetry, forbidding the existence of a Noether current. This obstruction extends the non existence of currents for charged classes to the case of discrete symmetries.

\subsection{Structure of the non local operators in the buffer zone}

We now analyze the non local classes that might appear in the buffer zone $Z$. These are the classes of ${\cal A}_{\rm max}(Z)$. First, a non local class in ${\cal A}_{\rm max}(Z)$ may correspond to operators that are additive in $R\cup Z$. In this case the operator belongs to
\be
{\cal A}^{R}(Z)={\cal A}(R\cup Z)\cap {\cal A}(R)'\,.\label{def1}
\ee
Analogously, if the non local operator is additive in $\bar{R}\cup Z$ it belongs to
\be
{\cal A}^{\bar{R}}(Z)={\cal A}(\bar{R}\cup Z)\cap {\cal A}(\bar{R})'\,.\label{def2}
\ee
It is simple to verify that both ${\cal A}^{R}(Z)$ and ${\cal A}^{\bar{R}}(Z)$ are contained in $ {\cal A}_{\rm max}(Z)$. The reason is that they commute with $ {\cal A}(R)\vee  {\cal A}(\bar{R})$. This implies
\be
{\cal A}^{R}(Z)\vee {\cal A}^{\bar{R}}(Z)\subseteq {\cal A}_{\rm max}(Z)\;.
\ee
The opposite inclusion, namely ${\cal A}_{\rm max}(Z)\subseteq {\cal A}^{R}(Z)\vee {\cal A}^{\bar{R}}(Z)$, after taking commutants, is equivalent to
\be
 ({\cal A}_{\rm max}(R)\vee {\cal A}(\bar{R})) \cap ({\cal A}_{\rm max}(\bar{R})\vee {\cal A}(R))\subseteq {\cal A}(R)\vee {\cal A}(R)\,.
\ee
This follows by expanding in non local operators the two algebras intersected on the left hand side and using that ${\cal A}_{\rm max}(R)\cap {\cal A}_{\rm max}(\bar{R})=\{1\}$. See section \ref{dos}.  Therefore we have that
\be 
{\cal A}_{\rm max}(Z)={\cal A}^{R}(Z)\vee {\cal A}^{\bar{R}}(Z)\;.
\ee
The non local operators of $Z$ are then generated by the ones that are additive in $R\cup Z$ and the ones that are additive in $\bar{R}\cup Z$. Let us call generically these non local operators $c$ and $\bar{c}$ respectively. If we call $a$ to the non local operators in ${\cal A}_{\rm max}(R)$ and $b$ to the ones in ${\cal A}_{\rm max}(\bar{R})$, we have that $[a,\bar{c}]=[b,c]=0$. Since 
\be\label{def3}
{\cal A}^{R\tilde{R}}(Z)={\cal A}^{R}(Z)\cap {\cal A}^{\bar{R}}(Z)={\cal A}(R\cup Z)\cap {\cal A}(\bar{R}\cup Z)\;,
\ee
 may contain non local operators of $Z$, some of the non local operators in $Z$ may be additive on both sides.  

 The classes $c,\bar{c}$ commute because we can always split $Z$ in two pieces, one contiguous to $R$ and the other contiguous to $\bar{R}$, and choose representatives of each of the two types of classes localized in each of the pieces, and hence  additive in complementary regions. Therefore a generic operator in ${\cal A}_{\rm max}(Z)$ has an expansion
\be  
\sum_{\lambda\beta s} O_{\lambda\beta s} \, c_\lambda\, \bar{c}_\beta\,\tilde{O}_{\lambda\beta s}\,, \label{sop}
\ee
for $O_{\lambda\beta s},\tilde{O}_{\lambda\beta s}$ local operators in $Z$, i.e belonging to ${\cal A}(Z)$. The classes $c$ and $\bar{c}$ are by definition non local classes in ${\cal A}_{\rm max}(Z)$. However, an operator $c \,\bar{c}$ may be decomposable into several classes of ${\cal A}_{\rm max}(Z)$ by the action of local operators in $Z$, so that $c\, \bar{c}$ may not represent a unique irreducible class.

We note in passing that we have not used that the buffer zone is thin. This result is completely general and of a topological nature: if $R$ and $\bar{R}$ are disjoint, the non local classes of $Z=(R\cup \bar{R})'$ are generated by products of classes additive in $R\cup Z$ and classes additive in $\bar{R}\cup Z$, and we have the general expansion (\ref{sop}) for the maximal algebra of $Z$.

\subsection{Truly additive twists are as good as complete additive ones}

By the definitions (\ref{def1}), (\ref{def2}) and (\ref{def3}), any global group element $U(g)$ leaves the algebras ${\cal A}_{\rm max}(Z)$, ${\cal A}^{R}(Z)$, ${\cal A}^{\bar{R}}(Z)$, ${\cal A}^{R\bar{R}}(Z)$, and ${\cal A}(Z)$ invariant.\footnote{To verify this we remind that conjugation with a unitary sends commutant algebras into commutant algebras.} It will mix the non local classes of the previous algebras between themselves if they are charged. However, a twist for $R$ generically only leaves ${\cal A}_{\rm max}(Z)$  invariant. In case the twist is additive, $\tau_g(R, Z)\in {\cal A}(R\cup Z)$, it leaves ${\cal A}^{R}(Z)$ invariant too. If the twist is complete  it leaves ${\cal A}^{\bar{R}}(Z)$ invariant, the reason being that the complementary twist is additive and the global symmetry group leaves the algebras invariant. Finally, if the twist is additive and complete, it additionally leaves  ${\cal A}^{R\bar{R}}(Z)$ invariant. 

The converse of these statements is not true in general. But it is true that if we have an additive twist leaving ${\cal A}^{R\bar{R}}(Z)$ invariant  we can construct another twist that is complete and additive. We call an additive twist keeping ${\cal A}^{R\bar{R}}(Z)$ in itself ``truly additive''. Complete additive twists are truly additive.  Below we show that the classes are invariant if and only if there is a complete additive twist, and the same is true for truly additive twists. 
 Conversely, if the classes are charged, it is not possible to have an additive  twist that keeps ${\cal A}^{R\bar{R}}(Z)$ in itself.

To prove this take a truly additive twist $\tau_g(R, Z)$ and an additive twist  $\tau_g(\bar{R}, Z)$. Using also the global symmetry transformation, we can define the following unitary
\be
z=U(g)^{-1}\, \tau_g(R,Z)\,\tau_g(\bar{R}, Z)\in {\cal A}_{\rm max}(Z)\,.\label{equis4}
\ee
We see that $O\, z= z \,\bar{C}$ for any $O\in  {\cal A}^{R\bar{R}}(Z)$ and some $\bar{C}\in {\cal A}^{\bar{R}}(Z)$.  It follows that 
\be
\sum_\lambda O_\lambda \,z\,  \tilde{O}_\lambda =z\, \bar{C}\,,\label{myyo}
\ee
for any $O_\lambda, \tilde{O}_\lambda \in {\cal A}^{R\bar{R}}(Z)$ and some $\bar{C}\in {\cal A}^{\bar{R}}(Z)$. Expanding $z$ in elements $c \bar{c}$ as in~(\ref{sop}), we can choose any of these elements by a left right action of elements in $O\in  {\cal A}^{R\bar{R}}(Z)$ (as in the left hand side of (\ref{myyo})). We thus get 
\be
C \, \bar{C}'= z\, \bar{C} \,.
\ee
In terms of the classes of ${\cal A}_{\rm max}(Z)$ with respect to the action of ${\cal A}^{\bar{R}}(Z)$, this equation implies there could be only a unique class $C$ in the expansion of $z$.  Since $z$ is unitary, the class $c$ must be Abelian, $c c^*$ is the identity class (here ${\cal A}^{\bar{R}}(Z)$), and we can take a representative $C$ unitary. All elements of the algebra ${\cal A}^{\bar{R}}(Z)$ in the expansion of $z$ can then be put on the same side, and we have
\be
z=  C\,\bar{C}\,,
\ee
with  $\bar{C}$ unitary in ${\cal A}^{\bar{R}}(Z)$.
  Then we get
\bea
1 &=& C^{-1}\, U(g)^{-1}\, \tau_g(R,Z)\,\tau_g(\bar{R}, Z)\, \bar{C}^{-1}= U(g)^{-1}\,((C')^{-1}
\, \tau_g(R,Z))\,(\tau_g(\bar{R}, Z)\,\bar{C}^{-1})\nonumber \\ &\equiv & U(g)^{-1}\,\tilde{\tau}_g(R,Z))\,\tilde{\tau}_g(\bar{R}, Z) \,,\label{sio}
\eea
with $(C')^{-1} \in {\cal A}^{R}(Z)$. Therefore (\ref{sio}) displays two additive complementary twists $\tilde{\tau}_g(R,Z)$ and $\tilde{\tau}_g(\bar{R}, Z)$ that are simultaneously additive and complete. 

Truly additive twists, and in particular complete additive twists, leave not only ${\cal A}^{R \bar{R}}(Z)$ invariant but also the additive algebra ${\cal A}(Z)$ and indeed all classes of ${\cal A}^{R \bar{R}}(Z)$. In fact any twist that leaves ${\cal A}^{R \bar{R}}(Z)$ invariant will do so. The reason is that classes of ${\cal A}^{R \bar{R}}(Z)$ with respect to ${\cal A}(Z)$ are dual to classes of ${\cal A}_{\rm max}(R\cup \bar{R})$ with respect to ${\cal A}_{\rm max}(R)\vee {\cal A}_{\rm max}(\bar{R})$. These dual operators are non local operators in $R\cup\bar{R}$ whose classes cannot be changed by non local operators in $R$ or in $\bar{R}$. Then by transportability the twists cannot change these classes. Therefore the twist cannot change the classes of ${\cal A}^{R \bar{R}}(Z)$ either. 

\subsection{Existence of additive complete twists for uncharged classes}
\label{exis}
In section \ref{types} we showed that the existence of complete additive twists implies the non local classes are uncharged. Now we prove the converse, if the classes are uncharged, there exist complete and additive twists. The proof is valid for internal global symmetries. We actually prove the standard twists are, at least, truly additive. Using the results of the previous section, this implies that we can always construct complete additive twists by slightly modifying the standard ones.

 Let us take an additive standard twist $\tau_g(R,Z)$ constructed with the split property. The complementary twist 
\be
\tau_g(\bar{R},Z)=\tau_g(R,Z)\,U(g)^{-1} \label{estano}
\ee
 gives a representation of the group. In what follows we will only need to consider an Abelian subgroup $G_g={g^n}$, with $n$ an integer, of $G$ generated by a single element $g$.   
 
The twist $\tau_g(\bar{R},Z)$ leaves invariant the elements of ${\cal A}(R)$. Since $U(g)$ leaves ${\cal A}_{\rm max}(R)$ invariant and $\tau_g(R,Z)$ acts only in the split factor ${\cal N}$, the complementary twist $\tau_g(\bar{R},Z)$ takes a non local operator $a\in {\cal A}_{\rm max}(R)$ into a non local operator $\tilde{a}$ that belongs to  
\be
{\cal A}_{\rm max}({\cal N})={\cal N}\vee \{a\}={\cal N}\vee {\cal A}_{\rm max}(R)\,.
\ee
Notice that ${\cal N}={\cal A}(R)\vee J \,{\cal A}(R)\, J$ and $J \,{\cal A}(R)\, J$ commutes with ${\cal A}(R)$ and ${\cal A}_{\rm max}(\bar{R})$, and therefore $J \,{\cal A}(R)\, J\subset {\cal A}^R(Z)$. Then ${\cal N}\subset {\cal A}(R\cup Z)$. It also follows that  ${\cal A}_{\rm max}({\cal N})$ has the same classes $a$ as ${\cal A}_{\rm max}(R)$ and commutes with ${\cal A}(\bar{R})$. It can then be thought as the maximal algebra associated with the factor ${\cal N}$.   

Let the classes of $R$ be uncharged under $G$. They are therefore uncharged under the Abelian subgroup $G_g$.  Then the twist $\tau_g(\bar{R},Z)$ leaves the classes of ${\cal A}_{\rm max}({\cal N})$ invariant because $U(g)$ does, and because $\tau_g(R,Z)$ is additive and cannot change classes. Taking the action of $G_g$ on a non local operator $a$ of $R$ (through conjugation) we can project into irreducible representations such that 
\be
 \tau_g(\bar{R},Z)\, a_p \,\tau_g(\bar{R},Z)^{-1}= e^{i\,\phi}\,  a_p\,.
\ee 
This $a_p$ belongs to the same class as $a$. As $\tau_g(\bar{R},Z)$ leaves elements of ${\cal N}$ pointwise invariant, and the elements of the class $[a]$ in ${\cal A}_{\rm max}({\cal N})$ are generated by $a_p$ and ${\cal N}$, we see that necessarily all elements in $[a]$ transform with the same phase factor. Therefore there is in fact only one irreducible representation in the decomposition of $a$ into irreducible representations of $G_g$. In the case of a non-compact group $G_g$ we should have to take $a_p$ projected to an interval of phase factors, but the interval can be as small as we want. At the end the result is the same and only a single phase and representations are involved.       

Therefore, from the expression (\ref{estano}) we have learned that the action of $U(g)$ and the additive twist $\tau_g(R,Z)$ on the non local element $a\in {\cal A}_{\rm max}(R)$ differs (at most) by a phase factor. In terms of charges in a continuum group, this means that the commutators of the global charge $Q$ and the local additive charge $Q_R^{\rm add}$ with $a$ differ (at most) by a term proportional to $a$,
\be
[Q,a]=[Q_R^{\rm add}, a ]+ \phi'\,a\,. \label{pirom}
\ee

Now we take an additive standard twist $\tau^{\rm add}_g(R,Z)$ for $R$ and a complete standard twist $\tau^{\rm com}_g(R,Z)$ for $R$. We remind that these two can always be found by properly choosing the split algebras. We then compute
\be
z= \tau_g^{\rm com}(R,Z)^{-1} \,\tau_g^{\rm add}(R,Z)\,. \label{esp}
\ee
This element belongs to ${\cal A}_{\rm max}(Z)$ because it commutes, by construction, with both ${\cal A}(R)$ and ${\cal A}(\bar{R})$. 
It acts on an element $a$ of  ${\cal A}_{\rm max}(R)$ by introducing a phase since $\tau_g^{\rm com}(R,Z)$ acts as the global group.  The important point is that it does not take $a$ out of  ${\cal A}_{\rm max}(R)$ to the buffer zone. The unitary $z$ can also be written using an expression analogous to (\ref{esp}) in terms of the complementary twists, that is, the complete and additive ones for $\bar{R}$. More concretely
\bea
z&=& \tau_g^{\rm com}(R,Z)^{-1} \,\tau_g^{\rm add}(R,Z)= (U(g)^{-1}\,\tau_g^{\rm add}(\bar{R},Z))^{-1}\,(U(g)^{-1}\,\tau_g^{\rm com}(\bar{R},Z) )\nonumber\\ &=&\tau_g^{\rm add}(\bar{R},Z)^{-1}\,\tau_g^{\rm com}(\bar{R},Z) \,. \label{esp1}
\eea
Hence, it also transforms $b$ elements by a phase factor. Now consider an element $O$ of ${\cal A}^{R\bar{R}}(Z)$. We remind that
\be
{\cal A}^{R\tilde{R}}(Z)={\cal A}^{R}(Z)\cap {\cal A}^{\bar{R}}(Z)={\cal A}(R\cup Z)\cap {\cal A}(\bar{R}\cup Z)\;,
\ee
so that the commutant is
\be
{\cal A}^{R\tilde{R}}(Z)'={\cal A}_{\rm max}(R)\vee {\cal A}_{\rm max}(\bar{R})\;.
\ee
Therefore if $O$ belongs to ${\cal A}^{R\bar{R}}(Z)$ it commutes with non-local operators $a$ and $b$ at both sides. Further, since $O$ commutes with $a,b$ it also follows that
\be
z \, O\, z*  
\ee
commutes with $a,b$. We conclude that
\be 
z \, O\, z*\in {\cal A}^{R\bar{R}}(Z)\,.
\ee
We now use this information to prove that $\tau_g^{\rm add}(R,Z)$ is truly additive. Writing
\be\label{consz}
z=\tau_g^{\rm add}(\bar{R},Z)\, U(g)^{-1}\, \tau_g^{\rm add}(R,Z)\,,
\ee
we see the action of $z$ on $O$ is composed by the action of three operators. The twist $\tau_g^{\rm add}(R,Z)$ could take us from $O$ to an operator with an exclusive class of ${\cal A}^{R}(Z)$, that is, a class that is non additive in $\bar{R}Z$. This is because $\tau_g^{\rm add}(R,Z)$ is additive in $R\cup Z$. The global group element $U(g)$ will not change this class. The additive twist operator in the complement $\tau_g^{\rm add}(\bar{R},Z)$ cannot change an exclusive class ${\cal A}^{R}(Z)$ because it is additive in $\bar{R}\cup Z$. Therefore the only way for $z$ to leave  ${\cal A}^{R\bar{R}}(Z)$ in itself, satisfying~(\ref{consz}), is that $\tau_g^{\rm add}(R,Z)$ also leaves this algebra invariant. Therefore the original additive standard twist $\tau_g^{\rm add}(R,Z)$ is actually a truly additive twist. Since truly additive twists can be converted into complete additive twists, as we showed in the previous section, this completes the proof of the existence of complete additive twists when the non local classes are uncharged under the global symmetry.

Summarizing, if the non local classes of a region $R$ are invariant under the symmetry, there exist complete and additive twists for $R$. Further they can be constructed in a standard manner. This implies that the complementary twist is also complete and additive. Therefore the classes of $\bar{R}$ are invariant. Dual classes are both charged or uncharged together.

For continuum spacetime symmetries additive and complete twists for complementary regions cannot be complementary, though the charges can add to the global one. But the main obstacle for the generalization of the present proof is that we need a complete twist for ${\cal A}_{\rm max}(R)$ that has the same action as the additive twist in ${\cal N}$. It would be interesting to study this problem further. 

\subsection{Twist concatenation. Concatenable additive twists}
\label{conca}

\begin{figure}[t]
\includegraphics[width=0.8\linewidth]{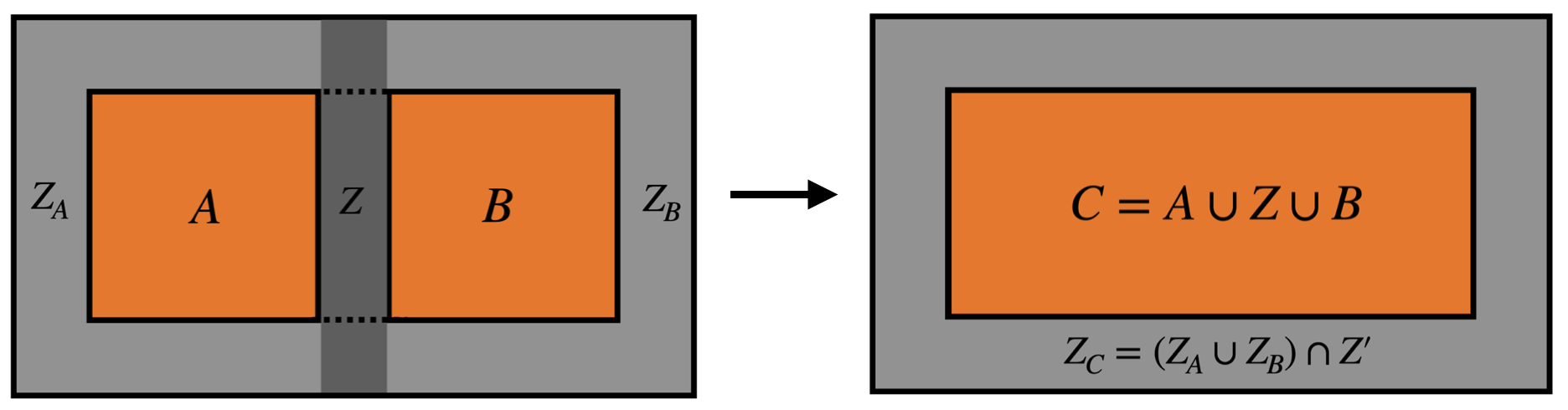}
\centering
\caption{Geometric configurations appropriate for the definition of concatenability of twist operators. The left hand side represents the existence of twist operators $\tau_g(A, Z_A)$ and $\tau_g(B, Z_B)$. The right hand side  represents the existence of a twist operator $\tau_g(C,Z_C)$ for $C=A\cup B\cup Z$ and $Z=(Z_A\cup Z_B)\cap Z'$. We say $\tau_g(A, Z_A)$ and $\tau_g(B, Z_B)$ are ``concatenable'' if $\tau_g(AZB,Z_A\cup Z_B)$ is the product of $\tau_g(A, Z_A)$ and $\tau_g(B, Z_B)$ . }
\label{FigConcat}
\end{figure}

Consider non intersecting regions $A$ and $B$ such that $Z_A$ is a buffer zone for $A$ and $Z_B$ is a buffer zone for $B$. We assume $Z_B\cap A=Z_A\cap B=\emptyset$. This geometric configuration is depicted schematically in the left hand side of Fig~(\ref{FigConcat}). In this scenario, we say the twists $\tau_g(A, Z_A)$ and $\tau_g(B, Z_B)$ 
  ``concatenate'' in the region $A\cup B\cup Z$,   if the product of the two gives a twist for $A\cup B\cup Z$ with buffer zone $(Z_A\cup Z_B)\cap Z'$. More concretely
\bea
&\tau_g(A, Z_A)\,\tau_g(B, Z_B)=\tau_g(C,Z_C)\,,&\\
&C=A\cup Z\cup B\,,\hspace{.7cm} Z_C=(Z_A\cup Z_B)\cap Z'\,.&
\eea
This is depicted in the right hand side of Fig~(\ref{FigConcat}). In this sense, complementary twists concatenate to the global symmetry operator.  

If we have two additive twists $\tau_g(A, Z_A)$, $\tau_g(B, Z_B)$ that concatenate to a complete twist $\tau_g(C,Z_C)$, it is immediate that the three twists are simultaneously complete and additive. First, the twist $\tau_g(C,Z_C)$ is also additive because it is the product of additive operators in $C\cup Z_C$. Now writing  
\be 
\tau_g(A, Z_A) =\tau_g(C,Z_C)\,\tau_g(B, Z_B)^{-1}\;,
\ee
we see the twist $\tau_g(A,Z_A)$ is complete since $\tau_g(C,Z_C)$ implements group operations in ${\cal A}_{\rm max}(A)\subseteq {\cal A}_{\rm max}(C)$, but   $\tau_g(B, Z_B)^{-1}$, being additive, commutes with ${\cal A}_{\rm max}(A)$. Analogously, the twist $\tau_g(B, Z_B)$ is necessarily complete. 

Hence, in this situation none of the non local classes of $A,B,C$ can be charged under the global symmetry.  A particularly useful case is when $C$ does not have non local classes, implying that $\tau_g(C,Z_C)$ is automatically complete. This case naturally appears when $C$ is the full space or a region with the topology of a ball, under the assumption of Haag duality.\footnote{For an orbifold of a global internal symmetry group that is spontaneously broken there are alternative algebras with single ball sectors. See section \ref{orbi} below.} In such scenario, the existence of additive twists for $A$ and $B$ that concatenate to a twist for $C$ imply the additive twists for $A$ and $B$ are also complete, and therefore the classes for $A$ and $B$ are uncharged. 

In this way we reach a simple but important conclusion. When the non local classes associated with certain generalized symmetry are charged under the global symmetry group, additive twists cannot concatenate to twists in a ball or the full space.  

For the case of continuous symmetries, the local twists $\tau_g(R,Z_R)$ can be replaced by the local charges $Q(R,Z_R)$ generating these twists by exponentiation. The charge $Q(R,Z_R)$ has the same commutator as the global charge for ${\cal A}(R)$, and commutes with $\cal A(\bar{R})$. The notion of twist concatenability can be phrased in terms of these charges.  Very simply, the local charges $Q(A,Z_A)$, $Q(B,Z_B)$
 in $A\cup Z_A$ and $B\cup Z_B$\footnote{The charges are actually affiliated to the additive algebras, as mentioned in the previous section.} are said to concatenate to the charge $Q(C,Z_C)$ in $C$ if
\be
Q(C,Z_C)=Q(A,Z_A)+Q(B,Z_B)\,.
\ee  

\begin{figure}[t]
\includegraphics[width=0.45\linewidth]{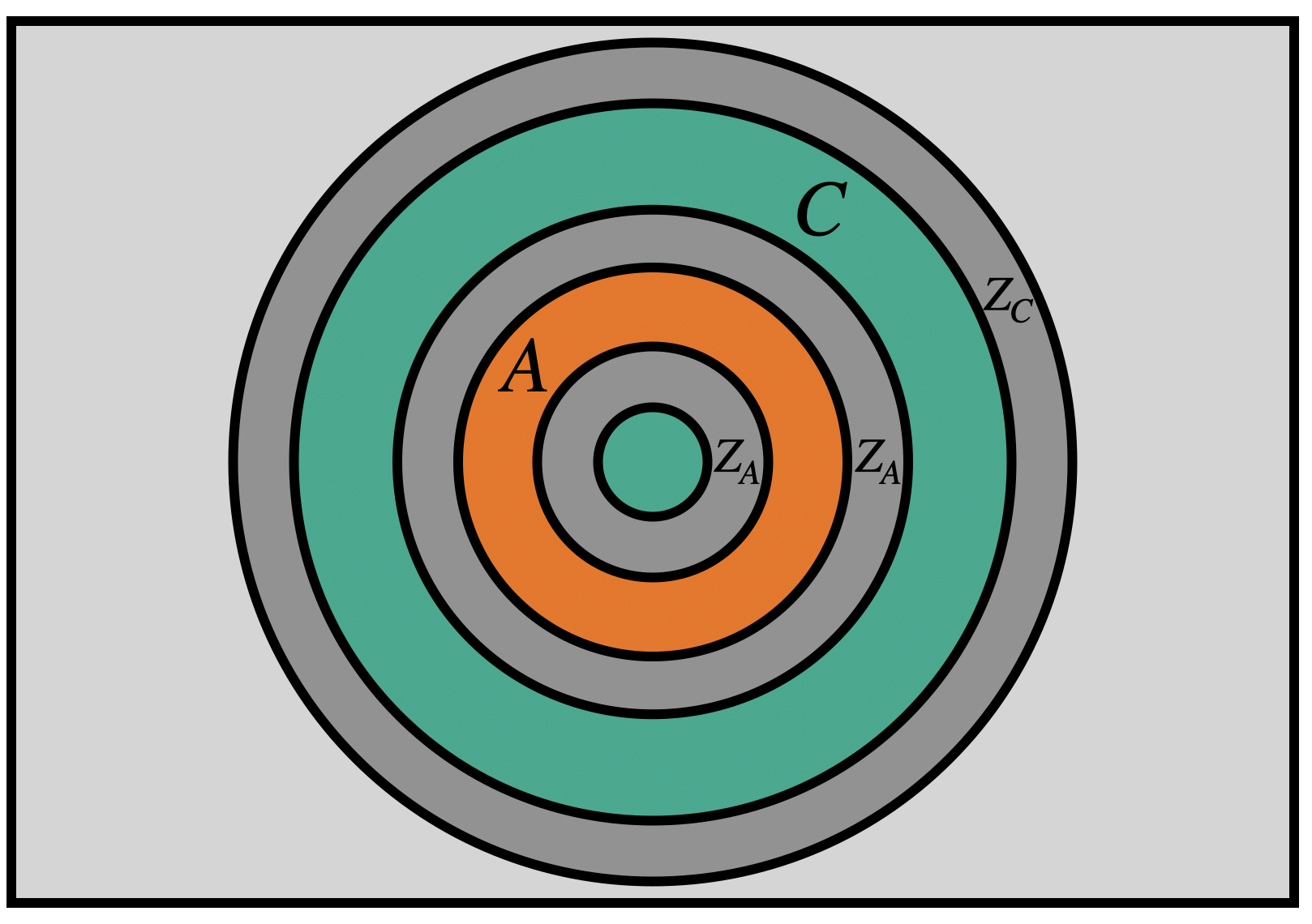}
\centering
\caption{A ball $B$ with its buffer zone $Z_B$. We have a region $R$ with buffer zone $Z$ inside it. Region $R$ might display generalized symmetries. When the classes associated with the generalized symmetry are charged under the global symmetry group additive twists in $R$ and $B-R$ cannot concatenate to twists in $B$.}
\label{FigConcatB}
\end{figure}

When $C$ does not have non local classes this again implies the non local classes in $A$ and $B$ are uncharged if the local charges of $A$ and $B$ are additive. The reason is that the commutator of $Q(C,Z_C)$ with a non local operator in ${\cal A}_{\rm max}(A)$ is equal to the commutator of $Q(A,Z_A)$ since $Q(B,Z_B)$ is additive. But since $Q(A,Z_A)$ is additive, these commutator cannot change the non local classes. Therefore  $Q(C,Z_C)$ and the global charge cannot change classes in $A$. This is the same argument used in section \ref{noetherz} for proving the non existence of Noether currents when there are charged classes.

\subsection{Arbitrary concatenation of twists} 
\label{arbitrary}

Given slightly separated regions $A_1,\cdots A_n$, allowing place for buffer zones $Z_1,\cdots Z_n$ for each one, we say that $\tau_g(A_1,Z_1),\cdots ,\tau_g(A_n,Z_n)$ are concatenable twists for $A_1,\cdots A_n$ with buffer zones $Z_1,\cdots Z_n$ if 
\be
\tau_g(A_1,Z_1)\cdots\tau_g(A_n,Z_n)=U(g)\,.
\ee
In this definition we also require $Z_i\cap A_j=\emptyset$, for any $ i,j$, and that $Z_i$ has a topology contractible to $\partial A_i$. Otherwise the idea of a concatenation of local operators is lost since the twist will be typically spread over all the associated buffer zone.

To construct such twists our first impulse would be to do it in order using the split property. We first divide $A_1$ with $Z_1$ from $\cup_{i=2}^n A_i$ and the remaining buffer algebras. More concretely we do the split between $A_1$ and $(A_1\cup Z_1)'$. Then inside the Hilbert space corresponding to $\cup_{i=2}^n A_i$ we split between $A_2$ and $\cup_{i=3}^n A_i$, etc. However, this construction will not respect the contractibility of the buffer zones. In most geometries, the buffer zone for $A_2$ will surround all $\cup_{i=3}^n A_i$ after the first split, etc. Then, if we split a plane with squares the typical twist in the partition will be totally non local in the plane. Similarly, if we do a split between $A_1$ and $\cup_{i=2}^n A_i$, and define the buffer zone to be $\cup_{i=1}^n Z_i$, then the twist of $A_1$ will be spread all over space.

We now consider the possibility of choosing a split partition ${\cal H}=\bigotimes_i {\cal H}_{{\cal N}_i}$ where the corresponding type I factors ${\cal N}_i$ contain ${\cal A}(A_i)$ and are included in the algebra ${\cal A}(A_i\cup Z_i)$. In such a case a twist concatenation would follow by restricting the action of the group to each factor, if the factors are chosen in a group invariant manner. 
However, such general splittability by local type I factors is in fact impossible for theories having generalized symmetries. For example, consider a split associated with squares, or other  topologically trivial regions.\footnote{See \cite{Huerta:2022cqw} for a manifestation of this phenomenon in the lattice.} The algebra $\vee_{i\in I}{\cal N}_i$ for any subset of indices $I$ is additive in the union of the corresponding regions. The complementary algebra $\vee_{i\notin I}{\cal N}_i$ is also additive. This implies none of these regions contains non local operators and there are no non local classes. In contrast, in an arbitrary  partition of the space by non intersecting regions it is  expected that the additive algebras corresponding to these regions generate the full algebra of operators (this is called strong additivity). However, these algebras are type III and are not in tensor product to each other because they share boundaries.
  
Then, we cannot expect arbitrary splittability by type I factors, at least when generalized symmetries are present. However, the question of the concatenation of twists remains. For Noether currents arbitrary concatenation of charges is possible, even if there are uncharged generalized symmetries. But arbitrary concatenation is not possible with charged generalized symmetries. Again, we think in using a partition by topologically trivial regions. If $\tau_k\,\tau_{k+1}\cdots\tau_{q-1}\,\tau_q$ covers a region with charged classes, it will be an additive twist for it, $U(g) (\tau_{q+1}\cdots\tau_n)^{-1}$ is complete, and $\tau_1\cdots \tau_{k-1}$ is additive. Then the classes cannot be charged, as explained in the previous section. Changing the ordering such that no contiguous subsequence covers a region having charged classes, we may obtain concatenability. But that means the commutators of the set of small additive twists must produce large non localities. 
 We see that when we concatenate more than two twists the issue of commutativity between the twists becomes important.     

Another question asks if we can produce concatenable twists when having uncharged classes and complete additive twists. We expect this to be the case but do not have a proof.  A particular case worth mentioning is the case in which the buffer zones $Z_{i}$ of $A_i$ share connected components between each other, that is, $Z_{ij}\equiv Z_{i}\cap Z_{j}$ is formed by unions of connected components of $Z_{i},Z_{j}$. In this case, as each connected component of the boundary divides the space in two disjoint regions, an adequate split partition can be defined, and we have that  the classes are uncharged if and only if there are commuting complete additive twists $\tau_{A_i}$ that concatenate.          

Suppose, more generally, that we have complete additive twists for $A$, $B$, and $C=A\cup B\cup Z$, $Z_C=(Z_A\cup Z_B)\cap Z'$, see Fig~(\ref{FigConcat}). Writing
\be
z=\tau_A \, \tau_B\, \tau_C^{-1}\,,\label{tomake}
\ee
we see the unitary $z$ carries ${\cal A}(Z_A\cup Z_B)$ into itself because each of the twists does. Then $z$ belongs to the additive algebra of the union of buffer zones.  This is not valid for other kind of twists, where $z$ typically contains non local operators of $Z_A\cup Z_B$. To correct the twists and achieve concatenability we need to split $z$ into a product of an operator in $Z_C$, to be absorbed into $\tau_C$, and another in $Z$, to be absorbed in $\tau_A$. This requires some locality of $z$. In the limit of small width $\epsilon$ of the buffer zone,  the scale of non locality, in the direction parallel to the boundary of $R$, of the split factors is expected to be of the same order $\epsilon$. In this limit we can think in a thin wall separating two half spaces,  
and in the UV there are no other scales in the problem than $\epsilon$. 
Hence, $z$, containing no non local operators, would be approximately local in the limit of small $\epsilon$ and the twists would  concatenate approximately in the limit. On the contrary, if $z$ contains non local operators, the limit $z\rightarrow 0$ cannot help with the concatenation.

\section{Examples}
\label{cinco}

In this section we describe several examples of generalized symmetries charged under a global symmetry. Examples will include continuous and discrete global symmetries, duality transformations, and spacetime symmetries. We will construct twist operators, and explicitly display some of their subtle properties we discussed more abstractly above.

\subsection{Orbifolds}
\label{orbi}
Given a theory ${\cal F}$ with an internal symmetry group $H$, we can produce another theory ${\cal O}={\cal F}/H$ formed by the invariant operators. This is called the orbifold theory. We want to see how the previous discussion reflects on this orbifold. The orbifold has generalized symmetries associated with the violation of duality in regions with the topology of two balls if the symmetry is unbroken or one ball if it is broken. In the unbroken case, the non local operators violating duality in a region $R$ consisting in two balls $R=B_1\cup B_2$ are intertwiners
\be 
I_r=\sum\limits_r\, \psi_r^i(x_1)\,(\psi_r^i)^{\dagger}(x_2)\,\,\,\,\,\,\,\,\,\,\,\,\,\, x_1\in B_1,\,\,\,\,x_2\in B_2\;.
\ee
These are charge anti-charge operators localized in the two balls. They are labeled by the representations $r$ of $H$. The operators violating duality in the complementary region $R'=(B_1\cup B_2)'$ are the symmetrized twists of $H$ acting on just one of the balls. These arise by taking the standard twists $\tau_h (B_1,Z)$ for $B_1$ described above, which are labeled by group elements $h\in H$, and symmetrized them by averaging over a conjugacy class $c$
\be 
\tau_c=\sum\limits_{h\in c}\, \tau_g\;.
\ee 
These invariant twists are then labeled by conjugacy classes $c$ of the group. In these cases, the original internal symmetry group $H$ acts trivially in the theory ${\cal O}={\cal F}/H$ by definition. It therefore acts trivially on each of the generalized symmetries of the orbifold, generated by the intertwiners $I_r$ and twists $\tau_c$. In fact the symmetry group $H$ itself does not exist any more in the orbifold since the global group operation is identified with the identity. For a more throughout treatment of these scenarios see \cite{Casini:2019kex,Casini:2020rgj,Review}.

We now build over these examples and introduce another internal symmetry group $G$  acting on  ${\cal O}$. We then have the alternative that $G$ leaves the classes of intertwiners and twists of ${\cal O}$ invariant or not. In particular, following the results of the previous section, if there is a Noether current for $G$, the first case must be realized. 

To see how the two situations might arise, imagine we have a bigger group of symmetries $\tilde{G}$ acting on ${\cal F}$ such that $H\subseteq \tilde{G}$ is a normal subgroup. Then $G=\tilde{G}/H$ is a group that acts on the orbifold ${\cal O}={\cal F}/H$. If $G$ is also normal in $\tilde{G}$, or equivalently $\tilde{G}=G\times H$, then the remaining global symmetry $G$ of the orbifold clearly does not interfere with the non local classes. However, if this is not the case, there will be twists of $G$ defined on two balls $B_1\cup B_2$ or on shells $S=(B_1\cup B_2)'$ acting non trivially on the non local classes. In this case, the twists $\tau_g$ associated with $G$ cannot be additive and complete at the same time, and additive twists cannot concatenate. If there are Noether currents associated with $G$ in the original theory ${\cal F}$, they will not belong to the orbifold ${\cal O}={\cal F}/H$.       

As an example, take a theory of two (possibly interacting) real scalar fields $\phi_1, \phi_2$, with an $H=Z_2\times Z_2$ symmetry $\phi_1\rightarrow - \phi_1$, $\phi_2\rightarrow - \phi_2$, plus a symmetry $\phi_1\leftrightarrow \phi_2$ of interchange of the two fields. We have
\be
\tilde{G}= (Z_2\times Z_2) \rtimes Z_2\equiv H\rtimes  Z_2  \;.
\ee
The last $Z_2$ in the semidirect product interchanges the two fields. It is a subgroup of the full symmetry group but not a normal one, while $H$ is indeed a normal subgroup.  Taking the quotient ${\cal O}={\cal F}/H$ we have an orbifold theory with a residual $G=\tilde{G}/H=Z_2$ symmetry of interchange of fields. We have four classes for two balls $B_1,\,B_2$, namely
\be 
1,\,\,\,\,\,\,\,\,\,\,\, I_1,\, \,\,\,\,\,\,\,\,\,\, I_2,\,   \,\,\,\,\,\,\,\,\,\, I_1\,I_2\;,
\ee
where $1$ is the identity class and $I_i=\phi_{i}(x_1) \,\phi_i(x_2)$ with $i=1,2$. The non local classes (the intertwiners) $I_1,\, I_2$, are interchanged by $G$, and therefore are charged under it. We also have four classes associated with the complementary region, which is a shell $S=(B_1\cup B_2)'$. These are
\be 
1,\,\,\,\,\,\,\,\,\,\,\,\, \tau^{H1}_{B1},\,\, \,\,\,\,\,\,\,\,\,\, \tau^{H2}_{B1},\,\,   \,\,\,\,\,\,\,\,\,\, \tau^{H1}_{B1}\,\tau^{H2}_{B1}\;,
\ee
where $1$ is the identity operator associated to the identity class, and $\tau^{Hi}_{B1}$ with $i=1,2$ are the localized twists, acting on $B_1$ as the global symmetry group $H$, and doing nothing in $B_2$. These twists are associated with the orbifold group $H$. Since $H$ is an Abelian group, such twists directly belong to the orbifold ${\cal O}={\cal F}/H$. They are also charged under $G$, as will be shown in a moment.

We now construct the localized twist of $G$ for a region $R$. This should be termed $\tau_R^G$, but since we will be using it repeatedly we rename it to $\tau_R^G\rightarrow \tau_R$. We note we can make twists for the transformation $ (\phi_1\rightarrow \phi_2, \phi_2\rightarrow -\phi_1)$ in the original ${\cal F}$ theory using the canonical commutation relations. To keep the discussion simple and not clutter the notation we omit the dependence on the buffer zone $Z$, and write the twist with the non smeared  heuristic form
\be
\tau_R= e^{i \frac{\pi}{2}\int_R d^{d-1}x\,(\phi_1\, \pi_2  -\phi_2 \pi_1)}\,.\label{exp}
\ee  
Here $\pi_1,\pi_2$ are the momentum variables. Changing the sign of the exponent will give $ (\phi_1\rightarrow -\phi_2, \phi_2\rightarrow \phi_1)$ . 

We need to see how these twists act in the orbifold theory ${\cal O}$. To start with, these twists interchange the two  algebras of neutral fields in ${\cal O}$,  
\be
\tau_R \,\phi_1(x_1)\phi_1(x_2) = \phi_2(x_1)\phi_2(x_2)\, \tau_R\,,\hspace{.7cm}\tau_R \,\phi_2(x_1)\phi_2(x_2) = \phi_1(x_1)\phi_1(x_2)\, \tau_R\,,\label{comm}
\ee
for $x_1,x_2\in R$. In the particular case in which $R=B_1\cup B_2$ is the union of two balls, this twist then interchanges the non local classes
\be 
\tau_R \, I_1=I_2 \,\tau_R \;.
\ee
However, (\ref{exp}) is not invariant under the independent change of sign for the two fields. Therefore, it does not belong to the orbifold. To construct the twist for $G$ on the orbifold we can average over the action of $H$. We can then redefine the twist as\footnote{To make this operator unitary we can divide it by $\sqrt{\cos^2\left( \frac{\pi}{2}\int_R (\phi_1\, \pi_2  -\phi_2 \pi_1)\right)}$.  This commutes with pairs of fields in the region.} 
\be
\tau_R=\frac{1}{2}\left( e^{i \frac{\pi}{2}\int_R (\phi_1\, \pi_2  -\phi_2 \pi_1)}+ e^{-i \frac{\pi}{2}\int_R (\phi_1\, \pi_2  -\phi_2 \pi_1)} \right)= \cos\left( \frac{\pi}{2}\int_R (\phi_1\, \pi_2  -\phi_2 \pi_1)\right)\label{ba}\,.
\ee
This twist is invariant under $H$. It has the correct commutation (\ref{comm}) on pairs of fields.
When $R$ is a ball, or any region with trivial topology, this operator is additive because its series expansion contains only pairs of fields of each type in $R$.

Instead, for two balls $R=B_1\cup B_2$ we have two choices. We can take the product $\tau_{B_1}\,\tau_{B_2}$ of twists for single balls. This is additive in the two balls by construction, but it will not be able to interchange the classes of the intertwiners, namely it does not send
 $I_1=\phi_1(x_1) \phi_1(x_2)$ to  $I_2=\phi_2(x_1) \phi_2(x_2)$. The reason is that these intertwiners contain a single field in each ball.  Complete twists for the interchange symmetry in the two balls have the same expression (\ref{ba}), where the integral is now on $R=B_1\cup B_2$. This is clearly non additive in the two balls, and involves the intertwiners themselves in its construction. This can be seen by expanding the cosine in a Taylor expansion. Also, if we have a ball $B$ containing $R=B_1\cup B_2$, then the additive twists of the form $\tau_{B_1}\,\tau_{B_2}$ and $\tau_{B-R}$, do not concatenate to the twist $\tau_{B}$. The reason is that $\tau_{B}$ transforms the non local classes in $R$, while $\tau_{B_1}\,\tau_{B_2}\, \tau_{B-R}$ does not. In contrast, the original twists in eq. (\ref{exp}), defined in the theory $\mathcal{F}$,  are concatenable. 

Consider now two nested balls, $B_1\subset B_2$, and the shell $S=B_2-B_1$. An additive twist $\tau_S$ for the shell is provided again by the formula (\ref{ba}) where $R=S$. This does not changes the non local classes given by the twists $\tau^{H1}_{B1},\tau^{H2}_{B1}$ of the orbifold with base on $S$. A complete twist $\bar{\tau}_S$ is provided by the complementary construction, by taking for example $\bar{\tau}_S\equiv\tau_{B_2}\tau_{B_1}^{-1}$ where both operators in the right are given by the additive expression (\ref{ba}). Since the group is $Z_2$ the inverse twist can be replaced by the twist itself. This gives an operator
\be
\bar{\tau}_S= \tau_{B_2}\,\tau_{B_1}\sim \tau_S + \cos\left(\frac{\pi}{2}\int_{B_1} (\phi_1\, \pi_2  -\phi_2 \pi_1)+\frac{\pi}{2}\int_{B_2} (\phi_1\, \pi_2  -\phi_2 \pi_1 )\right)\,.\label{34}
\ee
The first and second term on the right hand side have the same commutation (\ref{comm}) on local operators in $S$, while both terms commute with local operators of the orbifold in $B_1$. For the second term this is because the total coefficient of the integral of the current on $B_1$ is $\pi$ rather than $\pi/2$, what makes the transformation $\phi_1\rightarrow -\phi_1, \phi_2\rightarrow -\phi_2$, that leaves local operators on the orbifold unchanged. But this also shows the new operator in eq. (\ref{34}) contains, in addition to the additive twist $\tau_S$, the non local twists of the orbifold across $B_1$, which change the sign of the fields there. For a generic ball $B$, these were called $\tau_{B}^{H1}$ and $\tau_{B}^{H2}$. They change the sign of $\phi_1$ and $\phi_2$ in the ball $B$.\footnote{These can be written $e^{\pi \int_B \phi_i(x)\pi_i(x)}$.} If $B_1\subset B\subset B_2$, it is evident that
\be
\bar{\tau}_S\,\tau_{B}^{H1}=\tau_{B}^{H2}\,\bar{\tau}_S\;.
\ee
This says that the twist $\bar{\tau}_S$ is complete, and interchanges non local classes. The reason is that for two balls $B,\tilde{B}$ we have 
\bea
\tau_{\tilde{B}}  \tau_{B}^{H1}= \tau_{B}^{H2} \tau_{\tilde{B}}\,,\hspace{.7cm} \tau_{\tilde{B}}  \tau_{B}^{H2}= \tau_{B}^{H1} \tau_{\tilde{B}}\,,\hspace{.7cm}B\subset \tilde{B}\,,  \\
\tau_{\tilde{B}}  \tau_{B}^{H1}= \tau_{B}^{H1}\tau_{\tilde{B}}\,,\hspace{.7cm} \tau_{\tilde{B}}  \tau_{B}^{H2}= \tau_{B}^{H2}\tau_{\tilde{B}}\,,\hspace{.7cm}\tilde{B}\subset B\,.
\eea
Then, within $\bar{\tau}_S= \tau_{B_2}\,\tau_{B_1}$, the twist $\tau_{B_2}$ interchanges the orbifold twists while $\tau_{B_1}$, being a local operator inside $B$, commutes with them.

As an example where the global symmetry group $G$ is continuous, we consider the case of two massless real free scalar fields with symmetries $\phi_i\rightarrow \phi_i+\alpha_i$, $i=1,2$, and $G\equiv U(1)$ is the rotation between the fields. The full symmetry group is then isomorphic to the Euclidean symmetry of the plane
\be 
\tilde{G}=E(2)=T(2)\rtimes U(1)\;,
\ee
where the translations form the normal subgroup used to produce the orbifold $H=T(2)$, and where the residual global symmetry group is
\be
G=U(1)=E(2)/T(2)\;.
\ee
In this case the orbifold is generated by the derivatives $\partial \phi_i$. It has two-ball $B_1\cup B_2$ intertwiners of the form
\be \label{ints}
I_{\lambda\beta} =e^{i \,\lambda\, (\phi_1(x_1)-\phi_1(x_2))}\,e^{i \,\beta\, (\phi_2(x_1)-\phi_2(x_2))}\,\,\,\,\,\,\,\,\,\,\,  x_i\in B_i\;,
\ee
labeled by two real numbers $\lambda ,\beta$. The Noether current of the symmetry that rotates the fields is 
\be
J_\mu= \phi_1 \,\partial_\mu \phi_2-\phi_2 \,\partial_\mu \phi_1\,.       \label{curr}   
\ee
This is clearly not an operator of the orbifold theory since it contains fields $\phi_i$ that are not in the algebra of $\partial \phi_i$. This is in consonance with the generic results of the previous section, namely that non local operator labels charged under a continuous global symmetry forbid the existence of the Noether current. In this case, the complete twists for the remaining $U(1)$, constructed from the standard split construction transform $I_{\lambda\beta} \rightarrow I_{\beta\lambda} $. Notice these twists will belong to the orbifold, while the previous current does not.\footnote{The lack of the $U(1)$ current in the case of two dimensions was acknowledged in \cite{Current1} using other techniques. Here we see the physical origin of the missing Noether current, and how it fits into the general zoo of such QFT's.}

We have to be a bit more precise in this example, distinguishing the cases of dimension $d=2$ from those of $d>2$. For $d=2$ the fields $\phi_i$ are not quantum fields because of infrared divergences, while the orbifold theory generated by $\partial \phi_i$ still makes perfect sense. The rotation symmetry does not have a Noether current as already mentioned. This theory has two ball sectors generated by the intertwiners~(\ref{ints}). 

For $d>2$, the shift symmetry of the fields is spontaneously broken. Because of that, the algebra of the orbifold generated by $\partial \phi_i$ generates the full Hilbert space of the full free fields when acting on the vacuum. One then can take the view that the theory contains the scalars too, and there is a Noether current. But, consistently, from this perspective there are no non local sectors of a generalized symmetry associated with two balls, and there is nothing that forbids such a current. On the other hand, the orbifold algebra generated by $\partial \phi_i$ is algebraically closed. It is also closed under Poincare transformations. It therefore forms a subnet. This orbifold subnet, in addition to the two ball sectors due to the intertwiners, also has sectors on single balls. These are due to the local fields themselves. There is no Haag duality for single balls.  The commutant of the algebra of the derivative in one region contains the scalar field in the complement, and this scalar constitutes a non local operator in the orbifold. This is always the case for the orbifold algebras when there is SSB \cite{doplicher1990there,roberts1974spontaneously}. 

More precisely, the global symmetry group $G$ rotates between non local operators for single balls, but it is not generated by a Noether current in the orbifold. The single ball $B$ non local sectors are associated with operators of the form 
\be 
\psi^i_{\bar{\alpha}}=e^{i \int_B \alpha \, \phi_i}\,,\hspace{.7cm}i=1,2\,,\label{alpha}
\ee
where $\alpha$ is a smearing function with support inside the ball and $\bar{\alpha}=\int_B \alpha$. 
The non local class of $\psi^i$ depends exclusively of $\bar{\alpha}$. This is because an smearing function with zero integral gives an additive operator formed by $\partial \phi_i$.
The dual operators, naturally associated with the complementary region, are the orbifold twists 
\be 
\tau_\lambda^{Hi}= e^{i \lambda \int_{R} \dot{\phi}_i}\,,\hspace{.7cm}i=1,2\,.\label{lambda}
\ee
 The commutation relations between dual non local operators are
\be
\tau_\lambda^{Hi} \,\psi^i_{\bar{\alpha}}\,(\tau_\lambda^{Hi})^{-1}=e^{i\,\lambda\, \bar{\alpha}}\,\psi^i_{\bar{\alpha}}\,.
\ee
The twist for the global symmetry $G$ acting on the orbifold that is obtained from the current (\ref{curr}) reads
\be
\tau_\theta=e^{i \theta \int_B (\phi_1 \pi_2-\phi_2 \pi_1)}\,. \label{taur}
\ee
This twist is complete. It rotates the non local sectors. While it is an additive operator in the theory $\mathcal{F}$, it is clearly not an additive operator of the orbifold theory because it contains the fields $\phi_i$ themselves, consistent with the generic lessons of the previous section. To obtain an additive twist in the orbifold we first write
\be
\phi_i(x)=\phi_i(x_0)+\int_{x_0}^x dx^\mu\, \partial_\mu \phi_i
\ee
with $x_0$ a fixed point in $\partial B$. We then replace
\be
\phi_i(x)\rightarrow \int_{x_0}^x dx^\mu\, \partial_\mu \phi_i
\ee
in the expression (\ref{taur}). This rotates correctly $\partial_\mu \phi_i$, and therefore any additive operator in the orbifold, but it does not rotate the non local classes. It is not complete. The difference between the additive and complete twists is the operator  
\be
e^{i \theta \left(\phi_1(x_0) (\int_B \pi_2) -\phi_2(x_0) (\int_B \pi_1 )\right)}\,. 
\ee
This combines non local operators of the orbifold in $B$ (the fields $\phi_i(x_0)$) with non local operators of the orbifold based on the boundary region\footnote{More precisely these twists are non-local in the complementary region when include the necessary buffer zone for their definition. We are simplifying the discussion by taking the buffer zone small and thinking of these operators as living on the boundary of $B$.} $\partial B$ (the twist generators $\int_B \pi_i$). This is a general feature of complete twists that makes transparent their non-additivity.

We end this section with some remarks. First, other QFT's where the orbifold generalized symmetry is charged under a continuous symmetry  share similar features as the preceding example. In the first place, the symmetry giving place to the orbifold must be non compact,  as discussed in section \ref{uno}. Then, non compact global symmetries must be spontaneously broken in $d>2$ by the reconstruction theorem \cite{doplicher1990there}. By Goldstone theorem the theory is gapless.     

Second, the algebra of the derivative of a free massless scalar is also an example of a model that has sectors which are non invariant under conformal symmetries.  
 If we make a scale transformation the class labels cannot remain invariant because the class label $\bar{\alpha}$ in (\ref{alpha}) has dimensions $-(d-2)/2$ in energy, while $\lambda$ in (\ref{lambda}) has dimension $(d-2)/2$. Indeed, the orbifold algebra does not have a conformal stress tensor to generate the scale current because this requires improving (for $d>2$) and the use of the field $\phi$ in addition to the field derivatives. 

Finally, an analogous example can be constructed to produce ball sectors non invariant under Lorentz transformations. We take the subalgebra of derivatives of the free Maxwell field $\partial_\gamma F_{\mu\nu}$. The sectors with respect to this subnet are given by charged operators $e^{i \int_B \alpha_{\mu\nu} F^{\mu\nu}}$.  In \cite{buchholz1992new} this model was proposed as an example of spontaneous symmetry breaking of a continuous symmetry yielding a non scalar Goldstone boson (the photon here). Sectors are labeled by integrals $ \bar{\alpha}_{\mu\nu}=\int_B \alpha_{\mu\nu}$, and are not invariant under Lorentz transformations. Again, as expected, the stress tensor cannot be written in terms of the derivatives of $F_{\mu\nu}$. In other words, while the twists of the Lorentz symmetries can be constructed, to be complete they will necessarily involve the stress tensor of the full Maxwell theory, in particular the operator $F_{\mu\nu}$ that is a non local operator in the ball for the orbifold.

\subsection{Rotation symmetric Maxwell fields}

We now consider an example where the generalized symmetry classes are associated with the violation of duality in a ring like region. This arises with two independent Maxwell fields in $(3+1)$ spacetime dimensions
\be 
S=-\frac{1}{4} \int_{\mathcal{M}} d^4x \, F_{\mu\nu}^a F^{\mu\nu}_a \,,\quad F_{\mu\nu}^a = \partial_\mu A_\nu^a -  \partial_\nu A_\mu^a\,,\,\, \, a=1,2\,.
\ee
We have an Abelian gauge symmetry associated with the gauge transformations 
\be
 A_{\mu}^a \to A_\mu^a + \partial_\mu \alpha^a \,.
\ee 
This theory displays non local sectors for regions containing non contractible loops (ring like regions). The violation of duality in such regions corresponds to the flux of the electric or magnetic fields in surfaces bounded by these loops \cite{Casini:2020rgj,Review}. This is a non compact $1$-form generalized symmetry with group $R^2\times R^2$ (for $d=4$).\footnote{The two factors $R^2$ correspond to the two fields. For each field we have Wilson loops labeled by a real number and 't Hooft loops labeled by a real number as well, hence $R^2$.} 

In addition, this theory  also exhibits a global $U(1)$ symmetry corresponding to the rotation between the gauge fields. We want to analyze how this symmetry manifests itself in the gauge invariant algebra and Hilbert space. We immediately see this manifestation is going to be problematic by noticing that the infinitesimal transformations $\delta A_{\mu}^1 = \epsilon \,A_\mu^2$ and $\delta A_{\mu}^2= - \epsilon \,A_\mu^1$ of the $U(1)$ group are not implemented by a physical Noether current. The root of this feature is that the global symmetry transforms the non local sectors and the localized twists will be additive or complete but not both at the same time.\footnote{In this specific scenario, the non-existence of the Noether current is also a consequence of the Weinberg-Witten theorem \cite{WEINBERG198059}. This will be described in detail in the next section. An scaling argument against the existence of such current was given in \cite{Harlow:2018tng}.}

To analyze these features we start by computing the non-gauge invariant Noether current through the usual prescription. We obtain
 \be 
 J^\mu = F^{\mu\nu}_2 A_\nu^1 - F^{\mu\nu}_1 A_\nu^2 \,.
 \ee
This is conserved on-shell $\partial_\mu F^{\mu\nu}_a=0$.  For any region $R$ with buffer zone $Z$, this current gives candidate twists for the rotation symmetry as $\tau_q (R,Z) = e^{i \,q \, Q_R}$ where the local charge is
 \be 
  \quad Q_R= \int d^3x \, f(x)\, J^0(x) = \int_{R\cup Z} d^3x \, f(x)\, [E^i_1 (x)\, A^2_i(x) - E^i_2(x)\, A^1_i(x)]\,,
  \label{tuu}
  \ee
where $f(x)$ is a smearing function of compact support over the 3-dimensional region $R\cup Z$, with $f(x)=1$ for $x\in R$, and $f(x)=0$ for $x$ outside $R\cup Z$. In this scenario, since the theory is free, we only need to smear in space. This twist is not gauge invariant since the charge transforms as
  \be 
   Q_R \rightarrow Q_R - \int [\partial_i f(x)][E^i_1 (x)\, \alpha^2(x) - E^i_2(x)\, \alpha^1(x)] \,.
  \label{gauge}
  \ee
This is a smeared boundary term with support only on $Z$, because $\partial_i f(x)=0$ for $x\in R$.   The question now is if we can add some boundary term so as to make the charge gauge invariant. This question was considered before in ref. \cite{Harlow:2018tng}, where a (non smeared) version of the gauge invariant twist was constructed. Here we start by constructing an analogous smeared version. To make the smeared twist $\tau_q (R,Z) $ gauge invariant, consider the function $v(x,y)$ which is the solution of the Poisson equation with a unit charge at $x$ and Newman boundary conditions in the region $Z$
  \bea
  \nabla_y^2 \, v(x,y)=\delta(x-y)\,, \hspace{.7cm}x,y \in Z\;,
  \\ n^i\,\partial^y_i \, v(x,y)=0\,,\hspace{.7cm} y\in \partial Z \,. 
  \eea
where $n^i$ is the unit normal  to $\partial Z$. Using the function $v(x,y)$ we define the following non-gauge invariant quantity supported on $Z$
  \be 
  I_a (x) = -\int_Z d^3y\,  \partial^y_i v(x,y)\, A_i^a (y) \quad \to \quad \delta I_a(x) = \alpha_a(x)\,.
  \ee
Using it we further define the operator
  \be 
  C= \int d^3x\, \partial_i f(x)\,[E_i^1 (x)\, I_2(x) - E_i^2(x)\, I_1(x)]\,,
  \ee
that has the gauge transformation 
  \be
  \delta C =  \int d^3x\,\partial_i f(x) \, [E^i_1 (x)\, \alpha^2(x) - E^i_2(x)\, \alpha^1(x)] \,.
  \ee
This matches (\ref{gauge}) and the charge operator 
\be 
\tilde{Q}_R =Q_R+C\;, \label{toyo}
\ee
is gauge invariant. This is a smeared version of the charge constructed in \cite{Harlow:2018tng}. 

The charge $\tilde{Q}_R$ clearly implements the correct twist operation,  corresponding to the global $U(1)$ symmetry, in the additive algebra of $R$.  This is because the support of the added operator $C$ is confined to $Z$, hence the commutator with the electric and magnetic fields in $R$ coincides with the ones of the non invariant Noether charge $Q_R$.

For topologically trivial regions, this construction  gives an additive and complete twist since there are no non local sectors in these regions for this theory.  We now want to analyze this twist but defined for a ring $R$. We start by asking how this charge acts on non local operators in $R$. The non local electric operators have as generators
  \be 
 \Phi_a^E = \int d^3x\, {\Omega}_i (x)\, E^i_a(x)\,\,\,\,\,\,\,\,\,\,\,\,\, a=1,2\,,
  \ee
 where $\Omega$ is a smearing (vector) function whose curl defines a conserved current $ J =\nabla\times \Omega$ that has compact support on $R$.\footnote{See \cite{Pedro,casini2021generalized} for more detailed descriptions of the geometric setup and the connection with the associated 1-form symmetries.} Computing the commutator
 \be
[Q_R,\Phi^E_1]=  \int d^3x \,  \int d^3y\, f(x)\,{\Omega}^i(y)\,E^2_j (x) \, [A^1_j(x),    E^1_i(y)]=i \,\int d^3x \,  f(x)\,{\Omega}^i(x)\,E^2_i (x)   \,, 
\ee
we obtain a local operator in $R\cup Z$ (the support of $f(x)$). Using 
\be
[E_a^i(y),I^a(x) ]= \partial^i_y v(x,y)\,,
\ee
we also have
  \be
  [C,\Phi^E_1]
   = - \int d^3x\,\int d^3y\,{\Omega}_i (y) \,\partial_j  f(x)\,E^j_2 (x)\, \partial^i_y v(x,y)\,.
\ee 
This is an operator local in $Z$, and therefore also in $R\cup Z$. So the commutator of the twist with the electric flux is an additive operator in $R\cup Z$. This implies that this twist does not change the non local class of the electric flux.


 For the magentic flux we write 
 \be
 \Phi^B_a= \int d^3x \, {J}_i(x)\, A^i_a(x)\,\,\,\,\,\,\,\,\,\,\,\,\, a=1,2\,,
 \ee    
 with a conserved current $J$ with support in $R$. We have
\be
[\tilde{Q}_R,\Phi^B_a]= i \,\varepsilon_{ab}\, \Phi^B_b \,,
\ee
with $\varepsilon_{ab}$ the Levi-Civita tensor of two indices. 
This implies that these localized charges do rotate the non-local classes associated with magnetic fluxes in the same way as the global symmetry group.   

We conclude that this twist is neither complete nor additive. It produces the correct transformations in the additive algebra and in the non local magnetic fluxes, but it does not change the non local electric fluxes. Dualy, we can make a twist with the correct transformation for the non local electric fluxes and acting trivially on magnetic ones. In $d>4$ this same construction gives a complete twist for a ring $R$ which only contains Wilson loops as non local operators. This twist is non additive again. In this scenario, it is clear that additive twists cannot concatenate to the twist in a ball since the last one transforms all operators inside the ball.

\subsubsection{Additive twist by gauge fixing}

From the abstract discussion of the previous section we know on general grounds that additive and complete twists can be constructed. One way is to perform the split construction starting with the additive algebra in $R$ and the complete algebra in $\bar{R}=(R\cup Z)'$, where $Z$ is the buffer zone. This gives an additive twist for $R$ and a complete one for $\bar{R}$. Still it is clarifying to have a more explicit construction of the additive twist, and study the differences with the complete one.

To construct a purely additive twist in the present scenario we start with the expression (\ref{tuu}) and search for a solution of $\nabla \times A=B$, where the potential $A$ is written as a non local function of $B$ inside the region $R\cup Z$ only. This can be interpreted as a particular gauge fixing that takes into account the shape of the region, except for a small detail that we encounter below. Therefore, $A$, defined in this way as a functional of the magnetic field inside $R\cup Z$, is automatically gauge invariant and additive in $R\cup Z$, and so will be the twist constructed out of it. This type of gauge fixings where used in the context of evaluating the entanglement entropy for gauge fields. A general construction in the lattice is in \cite{Casini:2013rba}. In the continuum, and for the case of spheres it was discussed in \cite{Valentin}.  

\begin{figure}[t]
\includegraphics[width=0.7\linewidth]{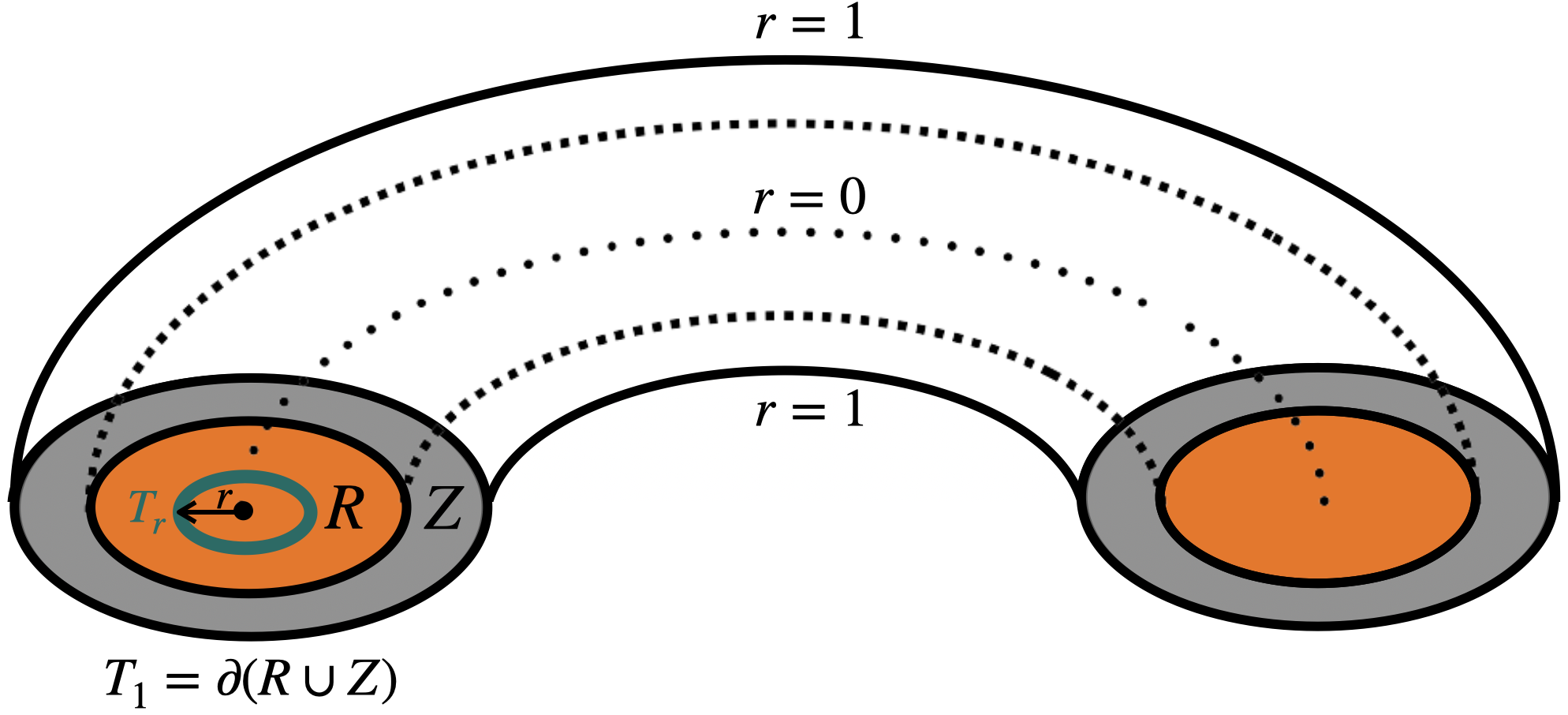}
\centering
\caption{Geometric configuration that facilitates solving for $\nabla \times A=B$ inside the ring. The surfaces of constant $r$ foliate the ring with topological torus $T_r$.}
\label{TorusM}
\end{figure}

To solve for $\nabla \times A=B$ in the ring $R\cup Z$, we take a ``radial'' coordinate $r\in [0,1]$ in $R\cup Z$. This radial coordinate is zero for certain non contractible loop inside $R$, and it is one at the surface $\partial \,(R\cup Z)$ of the ring. The surfaces of constant $r$ foliate the ring with topological torus $T_r$.  For $r=1$ we have $T_1=\partial \,(R\cup Z)$.   $\nabla r$ is perpendicular to the torus $T_r$. This geometric configuration is depicted in Fig~(\ref{TorusM}).

Let us call $F^{(r)}$ to the restriction of the two form $F_{ij}$ to the torus $T_r$.  Analogously, we call $A^{(r)}$ to the restriction of the one form $A_i$. On $T_r$ we have $F^{(r)}=d  A^{(r)}$, with $d$ the exterior derivative. By Hodge decomposition on $T_r$ we have 
\be
 A^{(r)}= d \alpha +\delta w +\bar{A}\,. \label{hodge}
\ee
The three terms are uniquely defined, and orthogonal in the sense of scalar product of fields on $T_r$. $\alpha$ is a $0$-form (a function), $w$ is a two form (characterized by a single function), $\delta$ is the adjoint of the exterior derivative $d$, and $\bar{A}$ is a harmonic one form (having zero Laplacian). $\bar{A}$ also satisfies $d\bar{A}=\delta \bar{A}=0$. We will use the gauge freedom to eliminate the first term in (\ref{hodge}). With this choice we have 
\be
F^{(r)}=d \delta w=(d\delta+\delta d) w=\Delta w\,,
\ee
with $\Delta$ the Laplacian operator. This equation has a unique solution for $w$ in terms of $F^{(r)}$, except for  additions of the volume two form. But this ambiguity does not affect $A^{(r)}$. Therefore, with this gauge choice we have uniquely fixed $A^{(r)}$ in terms of the magnetic field $F^{(r)}$ in $T_r$, except for the harmonic form $\bar{A}$.

To fix $\bar{A}$ we notice that harmonic forms are associated with the cohomology group of the torus and have a finite number of solutions in a compact manifold. In this case there are two solutions. They are locally pure gauge, but have non trivial circulation along the two non contractible circles of the torus. Therefore, once $\delta w$ is determined, their coefficient follows by evaluating the circulation of $A$, or the flux of the magnetic field, along any pair of surfaces bounded by the two non contractible circles in the torus. For the circle that is contractible inside $R\cup Z$, this is then given in terms of the magnetic field inside $R\cup Z$. For the other direction it cannot be computed in terms of operators in $R\cup Z$ alone. We need to fix the value of just one non-trivial circulation. We then fix this contribution to zero. This choice is not possible by gauge fixing $A$ in the full space, but it is a valid choice for solving the equation $\nabla\times A=B$ in $R\cup Z$.       

Therefore we have solved the tangential $A^{(r)}$ in terms of the magnetic field inside the ring. The radial component $A_r$ is fixed by the equation
\be
F_{a,r}= \partial_a A_r - \partial_r A_a\,,  
\ee
where $a=1,2$ are the two coordinates on the torus $T_r$. Since $A_a$ has already been fixed, this equation determines $A_r$ except for a constant value. This constant value can be set at will by the residual gauge transformation $A_r\rightarrow A_r + \partial_r f(r)$.  

With this choice of the vector potential $A$, the twist (\ref{tuu}) is automatically gauge invariant and additive. To verify it still does the right transformations on the additive algebra of the ring $R$, let us evaluate the commutator of
\be  
Q_R=\int d^3 x\,  f(x)\,  (E^1\cdot A^2-E^2\cdot A^1)\label{orig}
\ee
 with the electric and magnetic fields inside $R$. We start with the magnetic field
\be
[Q_R, B^a_i(y)]= i \, \varepsilon_{a,b}\, \int d^{3}x\,  \varepsilon_{ijk} \,\delta(x-y) \,\partial_k (f(x)\, A^b_j(x))=   i \,B^b_i(y)\,,
\ee
 where we have used 
\be
[E_i(x),B_j(y)]=-i \,\varepsilon_{ijk}\,\partial_k \, \delta(x-y)\,,  
\ee 
that the magnetic field commutes with the gauge fixed vector potential, and that $f$ is constant inside $R$. This demonstrates the correct transformation for the magnetic field. But we have not used our specific prescription for $A$ yet. To analyze the commutation with the electric field we  define $E=\nabla\times \tilde{A}$, and take for the expression of $\tilde{A}$ in terms of $E$ exactly the same one we have for $A$ in terms of $B$. Inserting $E=\nabla\times \tilde{A}$ in~(\ref{orig}) we obtain
\be  
Q_R=\int d^3 x\,  f(x)\,  (B^1\cdot\tilde{A}^2-B^2\cdot \tilde{A}^1)+ \int d^3x\, \varepsilon_{ijk}\, (\partial_j f)\, (\tilde{A}^1_k A_i^2- \tilde{A}^2_k A_i^1)\,. \label{last}
\ee
To simplify the last term in (\ref{last}) we choose our smearing function $f(r)$ a function of $r$. The last term can then be written as
\be
\int dr \, f'(r) \int_{T_r} (\tilde{A}^{(r)}_1\wedge A^{(r)}_2-\tilde{A}^{(r)}_2\wedge A^{(r)}_1)\,.
\ee
This vanishes since $\int_{T_r} \delta w\wedge \delta \tilde{w}=0$ and $\int_{T_r} \delta w\wedge \bar{\tilde{A}}=0$, by known properties of the calculus of forms on $T_r$, together with $\delta \bar{\tilde{A}}=0$.  Therefore
\be  
Q_R=\int d^3 x\,  f(r)\,  (B^1\cdot\tilde{A}^2-B^2\cdot \tilde{A}^1)\,, \label{lastik}
\ee
 has a dual invariant expression (see \ref{orig}). This induces the correct transformation on the electric field in $R$, replicating the previous calculation for the magnetic field. 

In conclusion, $Q_R$ with this particular gauge fixing  gives a generator of additive twists for $R$. These twists cannot change the non local classes since it is additive in both fields, but have the correct action on additive operators. The fully complete twist can now be constructed using the complementary twist of an additive one, namely
\be\label{dualtm}
\tau_{\textrm{complete}}(R,Z)=\tau_{\textrm{global}}\,\tau_{\textrm{additive}}(\bar{R},Z)^{-1}\;.
\ee

To see more clearly the difference between the complete and additive twists let us think heuristically in terms of sharp twists without smearing. The sharp additive charge $Q_{\rm add}$ rotates the electric and magnetic fields inside and at the boundary of $R$. It does nothing outside and therefore it does not rotate the non local operators, which are fluxes crossing the ring hole. To construct a complete charge out of  $Q_{\rm add}$, consider the electric and magnetic fluxes $\Phi^{1,2}_{E,B}(\Gamma_1)$ on the non contractible hole of $R$, and the ones $\Phi^{1,2}_{E,B}(\Gamma_2)$  on a contractible maximal circle of $R$. These commute, respectively, with local operators outside and inside the ring in the sharp approximation. The complete charge is then
\be
Q'=Q_{\rm add}+  (\Phi^1_{E}(\Gamma_1)\Phi^2_{B}(\Gamma_2)-\Phi^2_{E}(\Gamma_1)\Phi^1_{B}(\Gamma_2))+(\Phi^1_{B}(\Gamma_1)\Phi^2_{E}(\Gamma_2)-\Phi^2_{B}(\Gamma_1)\Phi^1_{E}(\Gamma_2))\;.
\ee  
This new charge rotates non local operators as well as additive ones. It is clear that this is a non local operator in $R\cup Z$. It also contains non local operators in the surface or buffer zone $Z$. It is also clear why two complete complementary twists cannot concatenate: the combination has twice the amount of non local operators needed. Two additive twists cannot concatenate either. A complete and an additive twist concatenate by construction, see~\ref{dualtm}. This is then another example of a QFT in which we can construct localized twists for any region. These are additive or complete, but not both at the same time, and therefore concatenability of additive twists (and complete twists as well) is lost.

\subsection{Duality symmetry}
\label{duality}
The electromagnetic duality symmetry of the Maxwell field 
\be
(F+ i F^*)\rightarrow e^{i\phi}\,(F+ i F^*)\,, \label{since}
\ee
is an internal symmetry of the theory since it preserves the local algebras. In particular it just exchanges the local electric field by the local magnetic field all over space. This implies that it interchanges 't Hooft and Wilson loops (electric and magnetic fluxes) and therefore it does not leave the non local classes associated with rings invariant. In accordance with the result of the previous section, there cannot be a physical Noether current for it.  Notice that eq. (\ref{since}) does not tell us how to implement the symmetry in the unphysical Lagrangian variables $A_\mu$. Therefore we cannot apply the usual prescription to write down a Noether current. Writing $F= d\wedge A$, $F^* = d\wedge \tilde A$, we have that 
\be
d\wedge (A\wedge F + \tilde{A}\wedge F^*)=F\wedge F+ F^*\wedge F^*=0\,.  
\ee
Hence the three-form $A\wedge F+\tilde{A}\wedge F^*$ is closed and its Hodge dual is a conserved current. In an explicit form we find
\be
J^\mu=\frac{1}{2}\epsilon^{\mu\nu\rho\sigma}\,( A_\nu\,\partial_\rho\,A_\sigma+\tilde{A}_\nu\,\partial_\rho\,\tilde{A}_\sigma)\,.
\ee
The corresponding charge is
\be
Q=\frac{1}{2}\int d^3x\, (A_i B_i + \tilde{A}_i E_i)\,.
\ee
It effects the desired transformations
\be
[Q,E_j]=i\,  B_j \,,\hspace{1cm} [Q,B_j]= -i \, E_j\,.
\ee
Therefore $Q$ implements the duality symmetry. From its implementation as the integral of a conserved density over all space it is a conventional (0-form) symmetry. But notice that such density is not gauge invariant. To construct an additive twist we proceed as above by writing $A$ and $\tilde{A}$ as a function of $B$ and $E$ respectively inside the ring like region $R$, obtaining
\be
Q_R=\int d^3x \, f(r)\, (A_i B_i + \tilde{A}_i E_i)\,.
\ee    
This  charge implements the duality symmetry correctly in the region $R$ by the same reasons. Being additive it cannot change the non local classes associated with 't Hooft and Wilson loops. To construct the complete twists one should use the complementary twists.

\subsection{Free graviton}
\label{gravi} 

The free massless graviton theory in $d=4$ is described by the curvature tensor $R_{(\mu\nu)(\alpha\beta)}$. This tensor is gauge invariant at the linearized level.  The theory is conformal invariant at the quantum level. For the split property in this case see \cite{longo2019split}. There should be local twists for the conformal group (spacetime symmetries). 
 One might ask if these twists are generated by local conserved currents.\footnote{In this particular case of $d=4$, the Weinberg-Witten theorem \cite{WEINBERG198059} applies, forbidding the existence of a stress tensor. We will talk in detail about such theorem in the next section.}

To answer this question we notice that Ref. \cite{casini2021generalized} showed that this theory has generalized symmetries originated by a $\mathbb{R}^{20}$ group of non local operators violating duality in ring like regions, namely regions with non contractible loops.\footnote{The generalized symmetries of the graviton field have also been recently considered in \cite{Hinterbichler:2022agn}.} These correspond to $1$-form generalized symmetries as in the Maxwell theory. Further, the non local classes were found to be charged under Poincare and conformal symmetries. In consequence, the results of the previous section forbid the existence of a stress tensor. This is because such putative stress tensor would give rise to additive, complete and concatenable twists, in contradiction with the fact that the non-local classes are charged under the conformal group.

In this section we display the generalized symmetries found in \cite{casini2021generalized} in a covariant form. This form is more suited for the analysis of the conformal transformations of the non local sectors, which we seek to do in detail. In particular in this example we want to clarify in a somewhat sophisticated example one of the observations described in the previous section, regarding the pointwise-like transformations associated with generalized symmetries charged under continuous groups.

We start by noticing that the on shell curvature of the graviton satisfies the relations
\bea
R_{(\mu\nu)(\alpha\beta)}&=& -R_{(\nu\mu)(\alpha\beta)}=R_{(\alpha\beta)(\mu\nu)}\,,\nonumber\\
\eta^{\mu\alpha}\, R_{(\mu\nu)(\alpha\beta)}&=&0\,,\nonumber\\
\varepsilon^{\mu\nu\alpha\delta}\, R_{(\mu\nu)(\alpha\beta)}&=&0\,,\nonumber\\
\varepsilon^{\mu\nu\gamma\delta}\,\partial_\gamma \,R_{(\mu\nu)(\alpha\beta)}&=&0\,,\nonumber\\
\partial^\mu \,R_{(\mu\nu)(\alpha\beta)}&=&0\,,\nonumber\\
\varepsilon^{\mu\nu}_{\,\,\,\,\,\,\gamma\delta}\,\varepsilon^{\alpha\beta}_{\,\,\,\,\,\,\sigma\epsilon} \,R_{(\mu\nu)(\alpha\beta)}&=& -R_{(\gamma\delta)(\sigma\delta)}\,.
\eea
The dual curvature tensor, which reads
\be
 R^*_{\gamma\delta\alpha\beta}= \frac{1}{2!}\,\varepsilon_{\mu\nu\gamma\delta} \,R^{(\mu\nu)}_{\,\,\,\,\,\,\,\,\,\,(\alpha\beta)}\,
\ee
satisfies the same relations.
 
The non local sectors for rings arise from the existence of closed two-forms $\omega$,  characterized by $d\,\omega=0$, constructed out of the curvature and that are not exact in the gauge invariant algebra. These forms can be integrated on a two dimensional surface 
\be
O_\Sigma=\int_\Sigma d\sigma\, \omega\,.
\ee
Since the form is closed, the integral does not depend on the particular two-dimensional surface provided that the boundary $\Gamma=\partial \Sigma$ is kept fixed. Therefore $O_\Sigma\equiv O_\Gamma$. In particular, $O_\Gamma$ commutes with all local operators localized spatially to $\Gamma$, because the surface $\Sigma$ can be displaced away from the local operator in that case. Hence, $ O_\Gamma$ will be a non local operator in any ring $R$ containing $\Gamma$ as a non-contractible loop.

Now, the spacetime region spatial to $\Gamma$ has  non contractible loops $\Gamma'$. If we fix $\Gamma'$ and continuously move $\Gamma$ to $\tilde{\Gamma}$ remaining in the causal complement of $\Gamma'$, the non local classes associated with $O_\Gamma$ and $O_{\tilde{\Gamma}}$ (with respect to $\Gamma'$) will be the same,  as far as we are still integrating the same $\omega$ in both surfaces, and $\omega$ is formed by local gauge invariant operators. The reason is that the difference between these two operators $O_\Gamma$ and $O_{\tilde{\Gamma}}$ is an integral of local operators in the complement of $\Gamma'$, that cannot change the class of $O_\Gamma$. We see in this case, as in any other where the non local sectors arise from integrating a gauge invariant local form, the notion of ``transportability'' described in the second section is automatic.     

Hence, to obtain the non local sectors, we need to find the set of independent two-form $\omega$'s satisfying
\be
\partial_\mu\, \omega_{\nu\sigma}+ \partial_\nu\, \omega_{\sigma\mu}+ \partial_\sigma\, \omega_{\mu\nu}=0\,,
\ee 
and such that $\omega$ is not $dA$ with $A$ gauge invariant. Equivalently, we need to find two forms $\omega^*$ (the Hodge dual of $\omega$) such that
\be
\partial^\mu\, \omega^*_{\mu\nu}=0\, .
\ee
It turns out that we can construct four families of independent two forms with zero divergence:
\bea 
A_{\mu\nu} &=& R_{(\mu\nu)(\alpha\beta)} \, a^{\alpha\beta}\,,\hspace{.7cm}  a^{\alpha\beta} = -a^{\beta\alpha}\,,\label{A}\nonumber\\
B_{\mu\nu} &=& R_{(\mu\nu)(\alpha\beta)}\,(x^\alpha b^\beta-x^\beta b^\alpha)\,,\label{B}\nonumber\\
C_{\mu\nu} &=& R_{(\mu\nu)(\alpha\beta)}\,c^{\alpha\beta\gamma}x_\gamma\,,\hspace{.7cm} c^{\alpha\beta\gamma}=-c^{\beta\alpha\gamma}=-c^{\alpha\gamma\beta}\,,\label{C}\nonumber\\
D_{\mu\nu} &=& R_{(\mu\nu)(\alpha\beta)}\,(x^\alpha d^{\beta\gamma}x^\gamma-x^\beta d^{\alpha\gamma}x^\gamma+\frac{1}{2} d^{\alpha\beta} x^2)\,,\hspace{.7cm} d^{\alpha\beta}=-d^{\beta\alpha}\,,\label{D}
\eea 
where the $(a^{\alpha\beta},b^\alpha,c^{\alpha\beta\gamma},d^{\alpha\beta})$ are real free parameters determining the specific two form. Given the symmetries of these free parameters, the total number of them is $6+4+4+6=20$. The fusion rules are simply the addition of the charges, giving rise to an Abelian $\mathbb{R}^{20}$ group of non local operators. No new independent two form is gained by using $R^*$ instead of $R$ in these expressions. To compare with \cite{casini2021generalized}, the family $A$ gives the operators called electric and magnetic translations in that work, the families $B$ and $C$ give the electric and magnetic dilatations and rotations, and finally the family $D$ gives the electric and magnetic special conformal transformations.

The  most general non local operator $O_\Gamma$  with parameters  $(a^{\alpha\beta},b^\alpha,c^{\alpha\beta\gamma},d^{\alpha\beta})$ in a ring like region $R$ containing $\Gamma$  can be written as
\be 
O_\Gamma=\int_\Sigma d\sigma^{\mu\nu}\,R_{\mu\nu\alpha\beta}(x) \, f^{\alpha\beta}(x)\,, \label{flux}
\ee 
where $d\sigma^{\mu\nu}= d x^\mu \wedge d x^\nu $ is the dual of the area differential two form of the surface ${\Sigma}$ with boundary $\Gamma$ and $f^{\alpha\beta}(x)$ represents all the functions defining the closed two forms  (\ref{A}). Its explicit form is
\be 
f^{\alpha\beta}(x)=a^{\alpha\beta}+(x^\alpha b^\beta-x^\beta b^\alpha)+ c^{\alpha\beta\gamma}x_\gamma+(x^\alpha d^{\beta\gamma}x^\gamma-x^\beta d^{\alpha\gamma}x^\gamma+\frac{1}{2} d^{\alpha\beta} x^2)\,.
\ee
 
Consider also a non local operator  $O_{\Gamma'}$ in the complementary region $R'$ with parameters $(\tilde{a}^{\alpha\beta},\tilde{b}^{\alpha},\tilde{c}^{\alpha\beta\gamma},\tilde{d}^{\alpha\beta})$. The operators $O_{\Gamma}$ and $O_{\Gamma'}$ are non locally generated in complementary regions and therefore do not commute with each other, as described in general terms in the introduction. In this specific scenario such commutator was computed in \cite{casini2021generalized}. It is given by 
\be
[O_\Gamma, O_{\Gamma'}]= i \, \left(\,a\cdot\cdot \tilde{d}^\ast+  \, 2\,b\cdot  \tilde{c}^\ast - \,2\, c^\ast \cdot  \tilde{b} - \, d^\ast\cdot\cdot\tilde{a}\right)\label{toic}
\ee 
where we are using the convention
\be 
d^*_{\alpha\beta} =\frac{1}{2!}\,\varepsilon_{\alpha\beta\gamma\delta}\,d^{\gamma \delta} \,, \quad c^*_{\alpha} =\frac{1}{3!}\,\varepsilon_{\alpha\beta\gamma\delta}\,c^{\beta \gamma \delta} \,,
\ee
Disregarding the precise numerical coefficients in~(\ref{toic}), the commutators must be conformally invariant, what practically fixes the structure of the result.

We can now analyze the effect of a general conformal transformation on a non local operator $O_\Gamma$. This follows from the associated coordinate change implemented by a Lorentz transformation and conformal factor on the curvature tensor. We can absorb the change of point by a coordinate transformation $x\to\tilde{x}$ on the integration over $\Sigma$. This transports the curve $\Gamma$ to $\tilde{\Gamma}$, but this transformation is irrelevant for the non local classes as long as it is small. As a result, the global spacetime conformal transformation can be written as a change of the parameters determining the non local class of the slightly displaced $\tilde{\Gamma}$. More concretely, the set of two-form polynomials multiplying $R_{(\mu\nu)(\alpha\beta)}$ in the above equations is mixed up under general conformal transformations.

\begin{figure}[t]
\includegraphics[width=8cm]{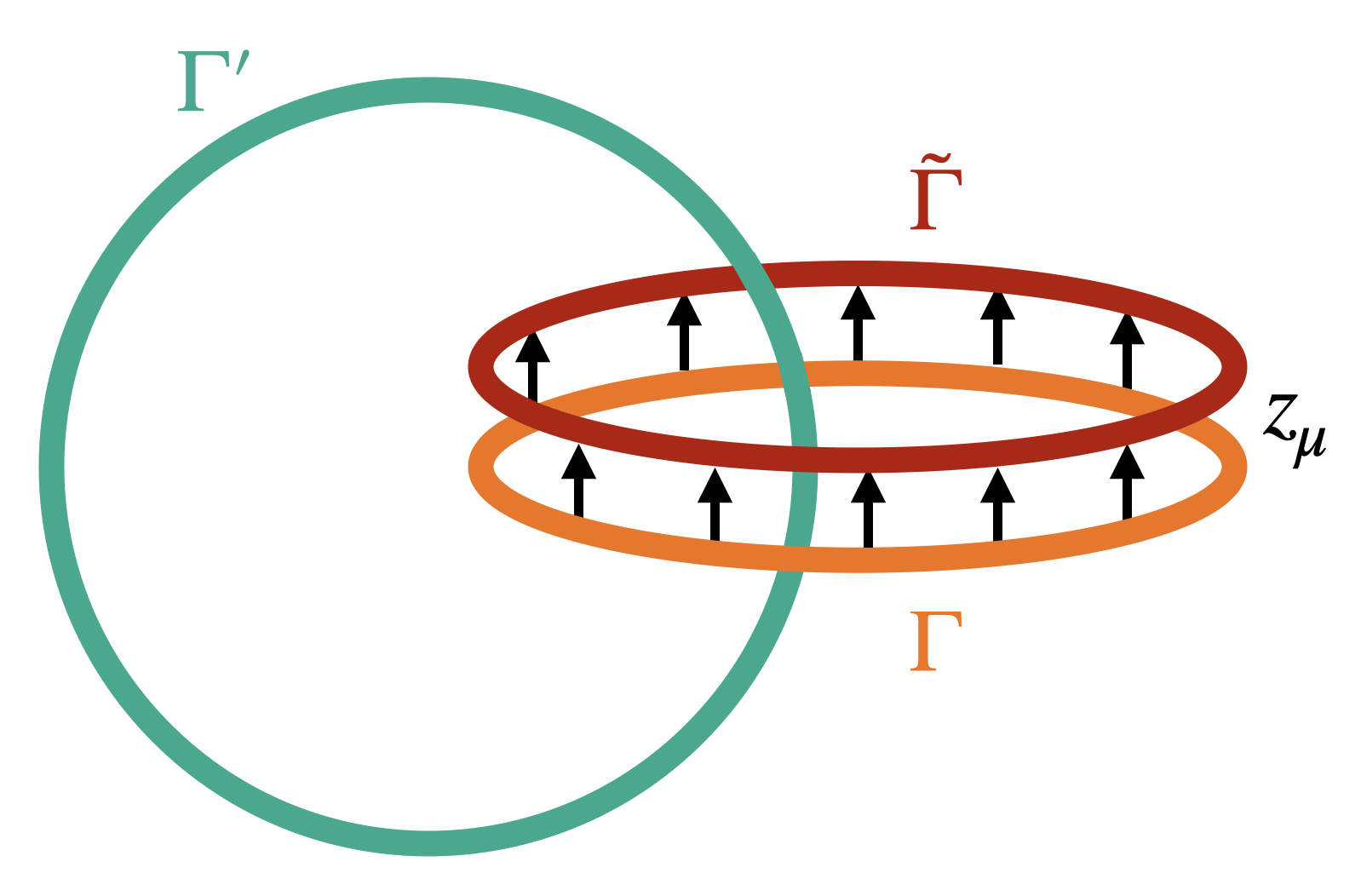}
\centering
\caption{The non local operator $O_\Gamma$ is defined with the curve $\Gamma$. In the complementary region we have $O_{\Gamma'}$, defined with $\Gamma'$, that does not commute with $O_\Gamma$. The curve $\tilde{\Gamma}$  is obtained when acting with small translations $z_\mu$ over $\Gamma$. The translation changes the non local classes for the graviton. However, to the operator of a specific class in $\tilde{\Gamma}$ it corresponds an operator with the same class in $\Gamma$ not by translating it, but by extending the surface integral from $\tilde{\Gamma}$ to $\Gamma$ with local operators. \label{notclass}}
\end{figure}

Let's start with translations. When performing $x^\mu \to x^\mu + z^\mu$, the Riemann tensor transforms as
\be 
U(z)\,R_{\mu\nu\alpha\beta}(x)\,U^{-1}(z) = R_{\mu\nu\alpha\beta}(x - z)\;.
\ee
This implies that the non local operator (\ref{flux}) changes as 
\be 
\tilde{O}_{\tilde{\Gamma}}=\int_\Sigma d\sigma^{\mu\nu}\,R_{\mu\nu\alpha\beta}(x - z) \, f^{\alpha\beta}(x) = \int_{\tilde{\Sigma}} d\tilde{\sigma}^{\mu\nu}\,R_{\mu\nu\alpha\beta}(\tilde{x}) \, f^{\alpha\beta}(\tilde{x}+z)\,.\label{kks}
\ee 
Note that $z^\mu$ is small enough to allow $\tilde{\Gamma}$ to have the same non local classes as $\Gamma$ (See Fig \ref{notclass}). By computing the change in the integrand in the right hand side of (\ref{kks}) we get the change of classes as
\bea
a^{\alpha\beta} &\rightarrow & a^{\alpha\beta}+(z^\alpha b^\beta-z^\alpha b^\beta)+ c^{\alpha\beta\gamma }z_\gamma+(z^\alpha d^{\beta\gamma}z^\gamma-z^\beta d^{\alpha\gamma}z^\gamma+\frac{1}{2} d^{\alpha\beta} z^2)\,,\nonumber\\
b^\alpha &\rightarrow & b^\alpha+d^{\alpha\gamma}z_\gamma \,,\nonumber\\  
c^{\alpha\beta\gamma} &\rightarrow & c^{\alpha\beta\gamma}+ (z^\alpha d^{\beta\gamma} +z^{\beta} d^{\gamma\alpha} + z^{\gamma}d^{\alpha\beta} )\,,\nonumber\\
d^{\alpha\beta} &\rightarrow  & d^{\alpha\beta}\,.
\eea
In an analogous way, one can verify that Lorentz transformations $x^\mu \to \Lambda^\mu_{\,\,\,\nu} \,x^\nu$  induce a transformation of  the class labels $(a^{\alpha\beta},b^{\alpha},c^{\alpha\beta\gamma},d^{\alpha\beta})$ simply as tensors with the inverse Lorentz transformation $\Lambda^{-1}$. 
For scale transformations we first notice that the Riemann tensor has scaling dimension three, so under scale transformations $x\rightarrow \lambda\, x$ it transforms as $
U(\lambda)\,R_{\mu\nu\alpha\beta}(x)\,U^{-1}(\lambda) =\lambda^{-3}\, R_{\mu\nu\alpha\beta}(\lambda^{-1} \,x)$. 
From this observation, we find the transformation of the class labels to be $a^{\alpha\beta}\rightarrow \lambda^{-1} \,a^{\alpha\beta}$ and $d^{\alpha\beta}\rightarrow \lambda \,d^{\alpha\beta}$, as expected on dimensional grounds. 

We are left with special conformal transformations 
\be 
x^\mu \to \frac{x^\mu + w^\mu\, x^2}{1+2\, w\cdot x +w^2 x^2}\,.
\ee
Such transformation acts locally as a dilatation and a Lorentz transformation. More precisely 
\be 
\frac{\partial x'^\mu}{\partial x^\nu} = \Omega (x) \, \Lambda^\mu_{\,\,\,\nu}(x)\,,
\label{local}
\ee
where $\Omega (x) $ characterizes a pure scale transformation given in terms of the vector $\omega$ as
\be 
 \Omega(x) = 1 - 2\, w\cdot x +w^2 x^2\;,
\ee
and $\Lambda^\mu_{\,\,\,\nu}(x)$ is a pure Lorentz transformation that reads
\be 
\left( \Lambda^{-1}\right) ^\nu_{\,\,\,\mu}(x)=\delta_\mu^\nu - 2\,\frac{w^\nu x_\mu -w_\mu x^\nu + w^2 x^\nu x_\mu +x^2 w^\nu w_\mu -2 (w\cdot x)w^\nu x_\mu}{1 - 2\, w\cdot x +w^2 x^2 }
\ee
Again we use that the curvature tensor on-shell behaves as a conformal primary with scaling dimension equal to three. It follows that
\be 
U(w)\,R_{\mu\nu\alpha\beta}(x)\,U^{-1}(w) =\Omega^{-3} \, \Lambda^{-1}{}^\rho_{\,\,\,\mu}\,\Lambda^{-1}{}^\sigma_{\,\,\,\nu}\,\Lambda^{-1}{}^\gamma_{\,\,\,\alpha}\,\Lambda^{-1}{}^\delta_{\,\,\,\beta}\,\,R_{\rho\sigma\gamma\delta}\left(\frac{x^\mu - w^\mu\, x^2}{1 - 2\, w\cdot x +w^2 x^2}\right)\,.
\ee
As above, we can change coordinates inside the flux. We then obtain a new non-local operator $\tilde{O}_{\tilde{\Gamma}} $ for the slightly displaced loop $\tilde{\Gamma}$ that reads
\be
\tilde{O}_{\tilde{\Gamma}}  = \int_{\tilde{\Sigma}\equiv\Sigma} d\tilde{\sigma}^{\mu\nu}\,R_{\mu\nu\gamma\delta}(\tilde{x}) \,\Omega^{-1}\left(x(\tilde{x})\right)\,\Lambda^{-1}{}^\gamma_{\,\,\,\alpha}\left(x(\tilde{x})\right)\,\Lambda^{-1}{}^\delta_{\,\,\,\beta}\left(x(\tilde{x})\right)\,f^{\alpha\beta}\left(x(\tilde{x})\right)\,.
\ee
We find that the classes labels transform under special conformal transformations as
\bea
a^{\alpha\beta} &\rightarrow & a^{\alpha\beta} \,,\nonumber\\
b^\alpha &\rightarrow & b^\alpha - 2\,a^{\alpha\gamma}w_\gamma \,,\nonumber\\  
c^{\alpha\beta\gamma} &\rightarrow & c^{\alpha\beta\gamma}+ 2\,(w^\alpha a^{\beta\gamma} +w^{\beta} a^{\gamma\alpha} + w^{\gamma}a^{\alpha\beta} )\,,\\
d^{\alpha\beta} &\rightarrow  & d^{\alpha\beta} + 2\,w^2 a^{\alpha\beta} + 4 \,( w^\alpha w_\gamma a^{\beta\gamma}-w^\beta w_\gamma a^{\alpha\gamma} ) - 2 \,(w^\alpha b^\beta -w^\beta b^\alpha  ) +2 \, c^{\alpha\beta\gamma}w_\gamma \nonumber\,.
\eea
We conclude that the action of the conformal group is linear in the charges. This is necessary to respect the fusion rules and keep invariant the commutators (\ref{toic}), in accordance with the general comments made in the second section.

In the Maxwell case we previously analyzed the electromagnatic duality symmetry from the present perspective, in which it is seen as conventional global symmetry that cannot have a Noether current given the results above. In the graviton case, in addition to the conformal group, we also have a $U(1)$ duality symmetry of the fusion rules and commutators. It corresponds to a rotation of $R$ and $R^*$
\bea 
\begin{pmatrix} R \\ R^\ast  \end{pmatrix} \to \begin{pmatrix} \cos{(\theta)}& -\sin{(\theta)} \\ \sin{(\theta)} & \cos{(\theta)} \end{pmatrix}\begin{pmatrix} R \\ R^\ast  \end{pmatrix}\,.
\eea
Using previous expressions we find that this transformations acts on the class labels as
\bea
a&\rightarrow &\cos(\theta)\, a-\sin(\theta)\,a^*\,,\hspace{.7cm} d^*\rightarrow \cos(\theta)\, d^*-\sin(\theta)\,d\,,\\
b&\rightarrow &\cos(\theta)\, b+\sin(\theta)\,c^*\,,\hspace{.7cm} c^*\rightarrow \cos(\theta)\, c^*-\sin(\theta)\,b\,.
\eea
Therefore, the full group of global symmetries acting on the non local sectors is the conformal group plus the duality group. Its Lie algebra is $SO(5,1)\times U(1)$, with $16$ generators. We have verified by explicit computation that the generic orbits of the full group of symmetries are $16$ dimensional. The $4$ Casimirs of the group $SO(5,1)\times U(1)$ must label the remaining $4$ coordinates of the class manifold. The quadratic one can be written as
\be 
a\cdot d + b\cdot b + c^\ast \cdot c^\ast \;.
\ee
However, as expected, there are special points in the manifold of class labels with non-trivial stabilizer groups. The dimension of these orbits is the dimension of the quotient of $SO(5,1)\times U(1)$ by the stabilizer group. All these results for the graviton case are in prefect agreement with the general discussion done in section~(\ref{pointl}).

The construction of twist operators that implement the conformal group transformations on the curvature tensor can be done, at the abstract level, by  the standard split construction, giving rise to additive or complete twists. More explicitly, for the case of spatial translations, a gauge non invariant twist can be constructed by using the $t^{0i}$ component of the Belifante  stress-energy tensor, or simply
\be 
\tau_P = e^{i z_j P^j}\,, \quad P_j = \int_R d^3x \, \pi^{kl}\partial_j h_{kl}\,.
\label{twistgrav}
\ee
The $h_{ij}$ are the graviton dynamical fields defined over a Cauchy slice that contains the region $R$, and $\pi_{ij}$ are the canonical momenta obeying
\be 
[h_{ij}(x), \pi^{kl}(y)]= \frac{i}{2}\left(\delta^k_i \delta^l_j + \delta^l_i \delta^k_j \right) \delta(x-y)\,.
\ee
The $P_j$ operator performs spatial translations inside $R$ over the canonical variables and the gauge invariant operators as expected.  But as could have been anticipated from the discussion of the Maxwell field, since the linearized diffeomorphisms act on the graviton dynamical fields as
\be 
\delta h_{ij}(x)= \partial_i \xi_j(x) + \partial_j \xi_i (x)\,, \quad \delta \pi^{ij}(x) = (\partial^i \partial^j - \delta^{ij} \partial^2 )\xi_0 (x)\;,
\ee
the proposed twist operator (\ref{twistgrav}) is not gauge invariant. However, as in the Maxwell case, its variation is simply the boundary term
\begin{align}
\delta P^j = \int_{\partial R} d\sigma_i &\left[ 2\partial_k \xi_0 \partial^j \partial^k \xi^i - 2 \partial^i \xi_0 \partial^j \partial_k \xi^k +\right.  2 \pi^{ik} \partial^j \xi_k  \\ 
& + \left. \xi_0 \partial^i \partial^j h^k_{\,\,k}- \partial^i\xi_0  \partial^j h^k_{\,\,k} - \xi_0 \partial_k \partial^j h^{ik} + \partial_k\xi_0  \partial^j h^{ik}\right] \nonumber\,.
\end{align}
Therefore, it should be possible to make this twist gauge invariant by adding a boundary term that cancels out the gauge transformation. If $R$ contains non contractible loops  this modification will transform the twist into a non local operator, consistent with the fact that there is not a well defined stress tensor for the graviton. Equivalently, this modification will contain the non local operators of the ring themselves in their construction. Also, one could approach the construction of the additive twist by a similar gauge fixing procedure, as done for the pair of Maxwell fields. We will not pursue this construction further.

\section{Generalizations of the Weinberg-Witten theorem} 
\label{ww}

One of the main results of this paper is that the existence of a generalized symmetry that is not invariant under a continuous global symmetry implies the global symmetry does not have a current. In this section we show how this result contains the Weinberg-Witten theorem \cite{WEINBERG198059}, and how it makes manifest the topological obstructions behind it. This new understanding will  allow several generalizations.

The Weinberg-Witten theorem \cite{WEINBERG198059} states that in four dimensional QFT's, massless particles of helicity $h \ge 1$  cannot carry a global charge associated with a conserved current, and that a QFT with a stress tensor does not admit massless particles of helicity $h >1$. The connection with the present results arises from two facts. First, free theories of massless particles with $h\ge 1$ in $d=4$ have $1$-form generalized symmetries (generalized symmetries associated with regions with non contractible loops). Second, for free theories of massless particles with $h\ge 3/2$ the non local operators associated with the one-form generalized symmetry carry Poincar\'e indices. Previously, we have analyzed explicitly the cases of $h=1$ and $h=2$. For higher integer helicity in $d=4$, we have the theory of a conserved field $R_{(\mu_1,\nu_1)\cdots (\mu_h,\nu_h)}$ with a $2\times h$ Young diagram symmetry for the indices. This field gives rise to different conserved two-forms in a way analogous to the Riemann tensor for $h=2$. Most importantly they are labeled by classes charged under Poincar\'e symmetry. Half integer spin also leads to conserved two-forms. For example, for $h=3/2$, the Rarita-Schwinger field $\psi_\alpha$ gives rise to a gauge invariant spinor two-form $\partial_\mu \psi_\nu-\partial_\nu \psi_\mu$. This two form generates ring sectors labeled by a constant spinor, and therefore charged under Poincar\'e symmetry.

Therefore, all these global symmetries, internal for $h\ge 1$ or Poincar\'e for $h> 1$, under which massless particles are charged cannot be generated by a current, as follows from our generic result, given the existence of $1$-form generalized symmetries in these theories that are charged under the global symmetry. This is the Weinberg-Witten theorem. In the present light, such theorem is rooted on the topological obstructions that are enforced when trying to charge a $1$-form symmetry with a $0$-form symmetry.

We remark we are not assuming neither the theory is free nor the theory has an exact $1$-form generalized symmetry. We are however assuming the global symmetry is exact, as in Weinberg Witten theorem. For example, the massless particles in Weinberg-Witten's theorem may not lead to an exact generalized symmetry in the full theory since there could be heavy charges breaking the non local operators at sufficiently high energies, where the theory becomes interacting. However, the asymptotic IR limit of the theory is free, as it is necessary for defining in and out states and ascertain that we have a massless particle in the spectrum. Therefore, in the IR these particles display the aforementioned generalized symmetries. On the other hand, the current, being associated with an exact global symmetry, has to generate the twists at all scales. Since the global symmetry is non trivially present in the IR theory, the current has to be part of the IR theory too. As we have seen, this is not possible.        

\subsection{Generalizations}

The first generalization that follows from the present approach concerns higher spin free massless fields in higher dimensions. Again, these masless particles cannot be charged under a continuous symmetry group implemented by a Noether current. This follows because these particles display
different types of generalized symmetries, such as $n$-form symmetries, and these cannot be charged under a symmetry generated by a local current.

We notice that massless particles in higher dimensions are characterized by representations of the semi-simple part $SO(d-1)$ of the little group. Fields are classified by representations of the Lorentz group $SO(d-1,1)$. The problem of classifying which massless particles can fit inside a given field type was recently solved in Ref. \cite{weinberg2020massless,Distler:2020fzr}. The solution is that a particle with a little group $SO(d-1)$ representation, characterized by a certain Young tableaux, can fit minimally into a field transforming under the Lorentz group $SO(d-1,1)$ in a irrep characterized by a Young tableaux given by adding another row (first row). Then, any non scalar bosonic particle will lead, at least, to a generalized $k$-form symmetry with $k$ the length of the largest column in the $SO(d-1)$ Young diagram. Global symmetries, under which these type of particles are charged, cannot have a current. If the $SO(d-1)$ Young diagram has more than one column, then the non local class labels will have Lorentz indices, and the theories describing these type of massless particles cannot have stress a tensor.       

The second generalization is that, in this language, we can get rid of any extra assumption about the spectrum of the theory, therefore reaching a statement valid for interacting QFT's or CFT's. Suppose a QFT contains a gauge invariant, closed, and ``physically'' non exact, $k$-form field $F$, with $1\le k\le d-2$. This means  $d F=0$, $F\neq d W$, for any $W$ gauge invariant $(k-1)$-form field.  In any such theory, if $F$ is charged under a continuous symmetry group $G$, there cannot be a current for $G$.  The reason is that such a closed, but non exact form leads to a $(d-k-1)$-form generalized symmetry. This is because by integrating $F$ on a $k$-dimensional open spatial surface gives rise to an operator that depends only on the $k-1$-dimensional boundary. Changing slightly the surface\footnote{As discussed in section \ref{gravi}, transportability is automatic in this scenario. }  of integration we can show it commutes with local operators spatially separated from the boundary. But this generalized flux is not an additive operator in the boundary since the closed form is not exact in the physical gauge invariant algebra.\footnote{It would be exact if $F=d W$ with $W$ a gauge invariant operator.} The charge of $F$ under the global symmetry leads to a charged generalized symmetry that does not admit a current as follows from our general result.

The case of $k=d-1$ ($0$-form symmetry) is a bit different because it leads to twists operators by integrating the $(d-1)$-form  over a $d-1$-dimensional spatial surface. This does not automatically leads to a generalized symmetry in the local QFT in the sense reviewed above. 
The reason is that the $d-1$ dimensional spatial surface of integration can only be moved in time. So it is not warranted that there are local operators uncharged under the twist. In other words, the orbifold generated by this symmetry may be simply empty if all operators are charged. We will face this issue again in the discussion section, when comparing with the Coleman Mandula theorem. In any case, if the orbifold is non trivial, and the twists are charged under a continuous global symmetry $G$ (possibly a spacetime symmetry), there cannot be Noether current in the orbifold theory generating $G$. 

As a particular case of these generalizations we can choose the continuous symmetry group to be the Poincare group itself. In this case, if the QFT has a gauge invariant tensor $v_{(s_1\cdots s_{k})\,\alpha_1\cdots \alpha_l}$, where $1\le k\le d-2$, which is closed for the first $k$ indices 
\be
\partial_{(s_{k+1}}   v_{(s_1\cdots s_{k}))\,\alpha_1\cdots \alpha_l}=0\,,
\ee
and it is not the exterior differential of a gauge invariant form $w_{(s_1\cdots s_{k+1})\,\alpha_1\cdots \alpha_l}$, then the QFT cannot have a stress tensor if $l\ge 1$. The indices $\alpha_i$ may be spinorial. 

More generally, any topological operator, not necessarily generated by a form field, of dimension $D$ in between\footnote{Here $D=0$ corresponds to intertwiners (not conventional local charged fields), $D=1$ to loop operators, etc.} $0\le D\le d-3$ charged under a continuous symmetry $G$ forbids a current for $G$. The reason is that the existence of a global current leads inevitably to the construction of concatenable additive twists for the  global symmetry acting on the generalized one.

Another generalization concerns discrete (not continuous) global symmetries $G$. Here, the non existence of a local current is replaced by the non existence of  additive concatenable twists. The constraints that might arise in these theories, due to this obstruction, are to be developed. 

The last generalization concerns non-relativistic QFT. As stressed in \cite{Review}, many of the tools used for the local approach to generalized symmetries are not restricted to relativistic QFT alone. In particular, we expect that on quite general grounds, generalized symmetries charged under continuous global symmetries forbid the existence of associated local conserved currents. If the global symmetry is discrete this condition will forbid the existence of concatenable twists. This result might be interesting for condensed matter systems. It also reduces the number of loopholes regarding the application of the Weinberg-Witten theorem in the context of quantum gravity. In this context, the Weinberg-Witten theorem is understood as an obstruction to having a purely QFT model of gravity, with an IR graviton. An obvious source of loopholes that has been exploited is the need of relativistic symmetry in the UV. Here we notice that even if the UV QFT is not relativistic it is highly constrained. In particular it cannot have currents generating any sort of spacetime symmetry, and it cannot be discrete with concatenable twists.

\section{Discussion}
\label{siete}

In this paper we have shown that global continuous symmetries that do not leave invariant some generalized symmetry cannot be generated by a local conserved current. Further, in these scenarios, the generalized symmetry must be non compact.\footnote{We remind that by ``non-compact'' we mean that the non-local classes and the dual non local classes both form a continuum.} One one hand, this provides a potentially complete characterization of the space of QFT's displaying  a violation of the strong version of Noether's theorem. On the other hand, this statement contains the Weinberg-Witten theorem as a special case, generalizing it in several non trivial directions.

These results have been derived from a careful analysis of the conditions for a generalized symmetry to be non invariant under a global one. First we noticed that global symmetries act as a point transformation in the space of non local classes. Second and most importantly, we developed a finer classification of twists operators in QFT. We found that twists can be additive (or not) and complete (or not), and that additive twists in smaller regions might not produce twists for larger ones. This suggests the concept of concatenability of twists, which we argue below to be crucial to understand Noether's theorem in QFT and aspects of global symmetries in quantum gravity.

We have supported the abstract discussion with several examples of generalized symmetries charged under a global one. When the global symmetry is continuous, hence without current and with a non compact generalized symmetry, always corresponded to free massless fields. We also analyzed cases with discrete global symmetries and electromagnetic duality.

In this final section we gather some conjectures and ideas that suggest themselves from the results of the paper, and clarify relations with previous known results.

\subsection{On the existence of currents for continuous symmetries}
\label{polo}

Having identified an obstacle for the existence of Noether currents associated with continuous symmetries, a natural idea is that currents should exist whenever this obstacle is absent. That is, the conjecture is that a theory with a continuous global symmetry, where all generalized symmetries for any topology are invariant under the symmetry, must contain a generating current for the symmetry. This is the converse of the main result present in this article. Equivalently, from a more mathematical perspective, in this article we have found a necessary condition for the existence of a Noether current. This necessary condition is the absence of generalized symmetries charged under the continuous global symmetry. The conjecture states that it is also a sufficient condition. 

In fact there are strong arguments in favour of this conjecture. It has remained a difficult technical task to derive quantum fields from a theory described in terms of local algebras. However, one promising path in the present scenario follows from the fact that, as we proved above, if the non-local classes associated with any topology are invariant under the global symmetry, there are additive complete twists for them. These additive twists for two regions with uncharged classes should concatenate, at least in an approximate sense as discussed in section \ref{arbitrary}, in the limit of thin buffer zones, to additive and complete twists for the union, if this later also has uncharged classes.\footnote{Proving, in a rigorous and general manner, the twist concatenability for theories with generalized symmetries but uncharged under the global symmetry group is an important open problem.} This is not possible for charged classes. Now, twists for a continuous symmetry can be described by local charges $Q_R$. Additive twists have local charges formed with additive operators of the region. If we have the time slice axiom we can think heuristically that these charges are additive operators on a time slice. That the additive twists are concatenable for any shapes of the involved regions implies that for any given partition of space, it is always possible to choose a partition of local charges that add up to the global one. This seems well in the way to show the charge can be expressed as an integral of a local quantity. Basically, notice that the validity of the notion of concatenability makes this problem analogous to the way one defines continuous integrals out of discrete sums, but for operator algebras. We expect one can prove it by thinking in limits of correlation functions for increasingly smaller partitions of the charge, and use convergence in the weak topology of von Neumann algebras.

\subsection{Are non compact generalized symmetries trivial?}
\label{perto}

The second conjecture suggested by this paper is that a Poincare invariant theory with non compact generalized symmetries is necessarily free and massless. This conjecture only concerns generalized symmetries. We do not require them to be charged under a further global symmetry.

In the first place, this conjecture is suggested by our failure in the construction of interacting examples. But there are further indications. Let us take the simplest case of two ball sectors (orbifolds). A non compact group behind the orbifold will have a continuum of charged sectors. This implies the symmetry has to be broken. The reason is that an unbroken symmetry should have a precise representation for each mass state, leading to infinite degeneracies and divergent densities of states. This is why the DHR reconstruction theorem \cite{Doplicher:1990pn} gives a compact unbroken symmetry group. For a CFT in the cylinder, each discrete energy eigenstate should have a specific representation too, leading again to a divergent density of states. Then, if we have a broken symmetry group in $d>2$, this leads to Goldstone bosons.\footnote{For a free massive scalar field, the current $j_\mu=f\,\partial_\mu \phi -\phi\, \partial_\mu f $, where $f$ is a solution of the Klein Gordon equation,  generates non compact symmetries of the net. However, they do not lead to a Lorentz invariant orbifold nor to Goldstone bosons.}  
 If the theory is conformal it would be a free theory.\footnote{Note, the same reasoning above about states in the cylinder implies the non compact symmetry is not conformal invariant. We comment more about this issue in a separate section below.} Another point is that a non Abelian non compact group would have infinite dimensional unitary irreducible representations, what would lead to a failure of the split property. This would reduce the possibilities to $\mathbb{R}^n$.   

The absence of a gap may be derived from the necessity of symmetry breaking of higher form  generalized symmetry too. In fact, a conformal continuous $1$-form symmetry in $d=4$ has to be free \cite{hofman2019goldstone}. Of course, the most interesting case is where conformal symmetry is not assumed from the start, and the crucial input is that there are dual continuous generalized symmetries, as shown in \cite{Casini:2020rgj,Review}. 

This conjecture entails deep dynamical implications. It connects renormalizability of certain theories with a question about symmetries. In this sense it describes new forms of anomalies. 
 For example, to make the Maxwell theory interacting without breaking the dual $\mathbb{R}^2$ generalized symmetries we should not introduce charges. The conjecture precisely forbids chargeless electrodynamics, i.e., interacting theories having dual continuous generalized symmetries. Lagrangians including a $(F^2)^2$ term or a dipole interaction with a neutron field $F_{\mu\nu} \bar{\Psi} \sigma^{\mu\nu}\Psi$ are indeed non renormalizable. To make sense of them the theory should be completed containing charges. This would in turn convert the non compact generalized symmetry into a compact one. 
This question deserves further study and we hope to regain it in a future work.

If this conjecture is correct, the examples concerning generalized symmetries charged under continuous symmetries, which necessarily imply a non compact generalized symmetry,  should follow the quite trivial pattern of the free  models we have discussed. The zoo of theories violating the strong version of Noether's theorem would be highly constrained and comprised by massless free models.

Further, combining this conjecture with the previous one in \ref{polo} leads to strong consequences. Both conjectures together would imply that a theory with a continuous global symmetry, but without current, must have a non invariant non compact generalized symmetry, from which the existence of a massless free sector would follow. This strongly improves Weinberg-Witten theorem, because the absence of the current, forcing the existence of the generalized symmetry, and associated with the massless particle in the IR, actually extends from the IR to the UV, making the graviton and the fotons in question real free particles. This further reduces the number of loopholes regarding the application of the Weinberg-Witten theorem to quantum gravity, and strongly supports its interpretation as an obstruction to having a purely QFT model of gravity. Of course, this leaves out theories with Hagedorn transitions where the split property cannot hold below some distance in space. This is the minimal distance where the idea of a tensor product structure for spatially separated regions can retain its meaning.  

Another consequence is that in theories without a free sector there could be no proper subnets of the net generated by the stress tensor. In particular, the stress tensor would be inside the algebra generated by any quantum field. 

\subsection{Scale invariance versus conformal invariance}
It is part of the standard lore in QFT that scale invariance should imply conformal invariance. In fact, only few counterexamples are known, and they are constructed with free theories \cite{El-Showk:2011xbs}. There is also a proof for $d=2$ \cite{zamolodchikov1986irreversibility,polchinski1988scale}, but not complete proof is known for higher dimensions. 

Assume there is a stress tensor, as it is usually assumed in the discussion of this subject. Theories without a stress tensor should be quite peculiar, possibly free as we argued above. Then, the usual route in trying to show conformal invariance from scale invariance involves improving the stress tensor such as to make it traceless. However, we see there can be an important obstacle in this construction. The theory with scale invariance may have generalized symmetry sectors with dimension-full class labels. In that case there is no possible dilatation current, and hence no improving of the stress tensor. Again, this type of counterexamples should be very peculiar because they have non compact generalized symmetries, and they are all probably free. We have seen the  example of the derivative of a free massless scalar in section \ref{orbi}. 
 Another one is the Maxwell field for $d\neq 4$ \cite{El-Showk:2011xbs}. This coincides with the derivative of the scalar in $d=3$. The reason of the breaking of conformal invariance is clear. By dimensional reasons the charges of 't Hooft and Wilson loops are dimension-full in these spacetime dimensions. So there is a stress tensor (generating Poincar\'e symmetries) but no dilatation current and therefore no ``CFT stress tensor''. 
 Notice that, as remarked in \cite{El-Showk:2011xbs}, for $d=3$ this problem can be somehow repaired by considering the full theory of a scalar instead of the derivatives of the scalar (that coincide with the Maxwell field algebra). The reason is that the sectors in this theory are orbifold sectors (two ball sectors) and can be eliminated by thinking in the mother theory reconstructed from the orbifold. The original model is a sub-theory of another one which is conformal invariant. However, this is not possible with the one form sectors in $d\ge 5$ for example. In  \cite{dymarsky2015scale} it was shown that if the theory with stress tensor and scale but not conformal invariance can be embedded in a CFT then it is a free theory. This is not the general case, but it agrees with our present discussion. We also remark that in the case there is no dilatation current nor stress tensor the model could still be conformal invariant, as happens with the graviton, or, more generally, with all primary free fields with Young tableaux index structure containing two rows or more. For these models, although the local currents do not exist for the reasons mentioned, the unitary representations of the conformal group do exist. On the other hand, if there is a stress tensor generating Poincar\'e symmetries, and no dilatation current, it is clear that conformal invariance is not possible without enlarging the model because the conformal transformations of the twists could be used to provide currents for all one parameter groups from any of the existing currents.     
 
  Note that for $d=2$, where an actual proof exists, all sectors are sectors of two (or more) intervals. That is, they are charge anti-charge operators, where the charge does not exist in the theory.\footnote{Here by ``charge'' we mean a generalized notion of it. It could be related to representations of the Virasoro and/or extended quiral algebras.} Sectors with dimension-full labels and scale invariance give a continuum of sectors, an example of the general description in section~(\ref{pointl}). The corresponding charge generator must be a conserved current, and this fixes its dimension to be $1$. In turn this gives no dimension-full sectors when the charge density is integrated over the interval. This is why this type of  counterexamples cannot exist in $d=2$. 

In conclusion, this gives a new understanding that separates the counterexamples from the rest of the QFT's, blaming the problem to the presence of sectors charged under the dilatation group for the dilatation invariant theory, implying no dilatation current and therefore no CFT stress tensor. 
Then, a natural conjecture is that the proof of the theorem should exist for the case where there is no such non scale invariant sectors. If we trust these cases only come from free theories, it would be enough to restrict the scope to models without a free sector. If we believe in the Noether theorem when there are no non invariant classes under the continuum symmetry, we would have a physical scale current plus the original stress tensor to start with, see \cite{Dymarsky:2013pqa}.

\subsection{Coleman-Mandula theorem}

The fact that a theory with a stress tensor cannot have generalized symmetries charged under Poincare symmetry resembles, and indeed extends in some directions, the Coleman-Mandula (CM) theorem \cite{coleman1967all}. In the directions it generalizes the CM theorem it also connects it with Weinberg Witten theorem  (see section \ref{ww}).  In other directions it does not. In this section we discuss the intersection of the statement in this paper with the CM theorem.

Recall that the CM theorem is expressed in the language of the $S$ matrix. It asserts that for an interacting theory with a mass gap, a group symmetry of the $S$ matrix must be a product of the Poincar\'e group and another group. The CM theorem does not assume a stress tensor. But, following our previous discussions, theories without stress tensor are probably free. Since the theorem does require interacting theories, considering theories with a stress tensor does not seem an important restriction. Though the theorem talks about $S$ matrix symmetries, to make a connection with the present paper we consider symmetries that act locally in spacetime.    

With the only assumption of having a stress tensor, the present results show that any generalized symmetry must be Poincare invariant. This extends the CM theorem in this precise direction, namely by applying it to higher form symmetries as well. However, the CM theorem is about $S$ matrix symmetries, and the only generalized symmetries that are (easily) visible in the $S$ matrix are the ones coming from global unbroken symmetry groups. In many cases these correspond to orbifold generalized symmetries ($0$-form symmetries), with Haag duality. Our statement applies to these orbifolds, and extends as well to broken symmetry orbifolds, forbidding sectors charged under Poincare symmetry. However, it requires there is a stress tensor in the orbifold itself. On the other hand, there may be also global symmetry groups that do not lead to an orbifold, because all local operators are charged. The orbifold is then empty. The cases where the orbifold does not have a stress tensor (in particular when there is no orbifold) are not covered by the present results.       

To understand this observations in more precise terms, we can form the additive algebra generated by the stress tensor. We call this algebra ${\cal T}$. For a CFT in $d=2$, this is the Virasoro net for the given central charge. In every dimension, the net ${\cal T}$ contains many local fields as well beyond the stress tensor and its derivatives. This theory ${\cal T}$ might have generalized symmetries. In particular it might have generalized symmetries associated with the violation of duality in two-ball regions. In this scenario, for $d>2$, the DHR reconstruction theorem \cite{doplicher1990there} asserts the existence of a global group $G$ of symmetries such that ${\cal T}$ is the orbifold over $G$ of an extended theory ${\cal F}$, namely ${\cal T}={\cal F}/G$. Following our results, since the stress tensor belongs to the orbifold theory, such global symmetry group $G$ must commute with Poincare symmetries to make the orbifold sectors uncharged. This restricted version of the CM theorem also follows from the DHR reconstruction itself by quite similar reasons \cite{buchholz1986noether}.

 However, this leaves open the possibility that there could be other symmetries that mix non trivially with Lorentz symmetry because the orbifold is empty or does not contain the stress tensor. This mixing of symmetries implies twist operators with indices. In the continuous case, these would be represented by conserved currents with more non antisymmetric indices\footnote{To compare with the discussion of the previous section \ref{ww}, this is the case $k=d-1$.} such as a symmetric $j_{\mu\cdots \nu}$, satisfying $\partial^\mu \,j_{\mu\cdots \nu}=0$. If we orbifold with these new symmetries implemented by the twists with Lorentz indices we get an algebra ${\cal O}$. This algebra might be empty, namely equal to the identity alone. If non-empty, the stress tensor must not belong to the orbifold $T_{\mu\nu} \notin {\cal O}$ since the orbifold sectors will display Lorentz indices and will be charged under the Poincar\'e group.\footnote{The case of ordinary internal symmetries uncharged under Poincare group has ${\cal T}\subseteq {\cal O}$, but this is not possible here.} Then we expect a free model since ${\cal O}$ does not have stress tensor. This is the case of the example discussed in section \ref{orbi}. We are then reduced to consider ${\cal O}=1$. We can also form the algebra generated by the current $j_{\mu\cdots \nu}$. This is equivalently the algebra generated by the twists of such symmetry. We can call it ${\cal J}$. If ${\cal J}\subset {\cal T}$, or more generally $T_{\mu\nu}\notin {\cal J}$, there is a net without stress tensor and we would again expect a free model.  
  Kac Moody algebras in $d=2$,  supersymmetry, and the conserved higher spin currents of  \cite{maldacena2013constraining} all have ${\cal O}=1$, ${\cal T}\subseteq {\cal J}$. Further ideas are needed to extend the present approach to constraint such scenarios. 
  
\subsection{Global symmetries in quantum gravity}  
  
In Ref \cite{Harlow:2018tng,Harlow:2020bee} an argument against the existence of conventional (0-form) symmetries in holographic theories of quantum gravity was provided. In between other aspects (of no direct relevance here), the argument requires decomposing a global symmetry group element $U(g)$, corresponding to a global QFT symmetry, into a product of twists for smaller regions. In  such references it was explicitly assumed that the ability of concatenating smaller twists to produce larger twists rests upon the validity of the split property of the QFT. In section~\ref{cuatro}, in particular in the last two subsections, we have seen that concatenability of local twists might or might not hold even if the split property holds for any region of spacetime, and twists can be defined. Splittability of the symmetry, in the sense used in \cite{Harlow:2018tng} of being able to construct local twists for any region of spacetime, is then not equivalent to arbitrary concatenability of the symmetry, in the sense of being able to decompose the global unitary into a product of local twists. The most direct physical reason for this inequivalence is that twists for larger regions contain in their definition operators that are non local from the perspective of the smaller regions and that cannot be created by simply multiplying operators in the smaller regions.

In this last discussion section we expand on the impact of this observation for the actual argument of Ref \cite{Harlow:2018tng,Harlow:2020bee}, and in more general terms for the folk theorem that states there are no global symmetries in quantum gravity \cite{Banks:2010zn}.

The non trivial requirement for the argument of \cite{Harlow:2018tng,Harlow:2020bee} to work is concatenability of localized twist operators. The QFT's in which we have explicitly shown that arbitrary concatenability cannot be achieved (even if localized twist operators exist for all regions) are QFT's with generalized symmetries charged under global symmetries. This result holds both for finite and continuous global symmetry groups. For continuous groups, as argued in the previous section, it might be the case that this scenario constraints the theory to be free. If this is the case it would not apply to holography. 

For finite groups the situation is more complicated. If we have a QFT with generalized symmetry charged under a finite group global symmetry we have no clue as to what constraints we can expect. This is an interesting avenue for future work. What it is clear is that in this case there are interacting examples. Crucial cases are gauge theories with electromagnetic duality, for example $\mathcal{N}=4$ SYM in $4d$. At the self dual point, electromagnetic duality can be seen as a global symmetry, as for the Maxwell case discussed in~(\ref{duality}). In these scenarios the non local operators associated with the generalized symmetry (Wilson and 't Hooft loops for example) are interchanged and therefore charged under duality transformations. If the generalized symmetry is discrete, as in $\mathcal{N}=4$ SYM with gauge group $SU(N)$, the duality symmetry is discrete. The QFT might be still interacting and such global duality symmetry is not concatenable. Quite interestingly, $S$-duality is also a symmetry of string theory, the dual to $\mathcal{N}=4$ SYM \cite{Maldacena:1997re}. Although the self dual point lies outside the holographic regime, we find very suggestive that $S$-duality, an exact symmetry of string theory and the dual QFT, is precisely a symmetry (at the self dual point) that escapes the argument of \cite{Harlow:2018tng,Harlow:2020bee}. It also suggests there might be other alike examples. We could have two $SU(2)$ gauge theories in the bulk, charged under a $Z_2$ symmetry interchanging them. The $Z_2$ generalized symmetries of each of the gauge theories are then charged under the global symmetry. In a putative boundary description this global symmetry should not be concatenable and again escape the wedge-reconstruction argument.  
On more heuristic grounds, if the black hole gauge charges of each gauge symmetry are conserved, the interchange symmetry might be protected. More generally, the intuition is that global symmetries might escape the folk theorem \cite{Banks:2010zn}, when they are protected by a gauge symmetry, the ultimate reason being the non-concatenability of dual twist operators.

Finally, for QFT's with generalized symmetries, with or without external global symmetry group, we have explicitly shown that one cannot make an arbitrary split of the QFT into many type I factors. As mentioned in section~\ref{arbitrary}, this does not necessarily imply the lack of twist concatenability, but this requires proof and indeed it might not be true. This also impacts holography since all known higher dimensional examples contain generalized symmetries, for example one form symmetries in the canonical $\mathcal{N}=4$ SYM scenario \cite{Maldacena:1997re,Aharony:2013hda,Hofman:2017vwr}.

\section*{Acknowledgements}
J.M. and H. C. thank many useful discussions with Stefan Hollands. The authors wish to thank also Marina Huerta, Edgar Shaghoulian and Gonzalo Torroba. This work was partially supported by CONICET, CNEA
and Universidad Nacional de Cuyo, Argentina.  The work of V. B.  is supported by CONICET, Argentina. The work of H. C. is partially supported by an It From Qubit grant by the Simons foundation. The work of J.M is supported by a DOE QuantISED grantDE-SC0020360 and the Simons Foundation It From Qubit collaboration (385592).

\bibliographystyle{utphys}
\bibliography{EE}

\end{document}